# Inverse magneto-refraction as a mechanism for laser modification of spin-spin exchange parameters and subsequent terahertz emission from iron oxides

R. V. Mikhaylovskiy[1,2a], E. Hendry[1], A. Secchi[2], J. H. Mentink[3], M. Eckstein[3], A. Wu[4], R. V. Pisarev[5], V. V. Kruglyak[1], M. I. Katsnelson[2], Th. Rasing[2] and A. V. Kimel[2]

[1]*School of Physics, University of Exeter, Stocker Road, Exeter, EX4 4QL, UK*

[2]*Radboud University Nijmegen, Institute for Molecules and Materials, Heyendaalseweg 135, 6525 AJ Nijmegen, the Netherlands*

[3]*University of Hamburg, Center for Free-Electron Laser Science, Luruper Chaussee 149, 22761 Hamburg, Germany*

[4]*Shanghai Institute of Ceramics, Chinese Academy of Sciences, Shanghai 200050, China*

[5]*Ioffe Physical-Technical Institute, Russian Academy of Sciences, 194021 St. Petersburg, Russia*

---

[a] Email for correspondence: R.Mikhaylovskiy@science.ru.nl

Ultrafast non-thermal manipulation of magnetization by light relies on either indirect coupling of the electric field component of the light with spins via spin-orbit interaction[1-23] or direct coupling between the magnetic field component and spins[4]. Here we propose a novel scenario for coupling between the electric field of light and spins via optical modification of the exchange interaction, one of the strongest quantum effects, the strength of which can reach $10^3$ Tesla. We demonstrate that this isotropic opto-magnetic effect, which can be called the inverse magneto-refraction, is allowed in a material of any symmetry. Its existence is corroborated by the experimental observation of THz emission by magnetic-dipole active spin resonances optically excited in a broad class of iron oxides with a canted spin configuration. From its strength we estimate that a sub-picosecond laser pulse with a moderate fluence of ~ 1 mJ/cm$^2$ acts as a pulsed effective magnetic field of 0.01 Tesla, arising from the optically perturbed balance between the exchange parameters. Our findings are supported by a low-energy theory for the microscopic magnetic interactions between non-equilibrium electrons subjected to an optical field which suggests a possibility to modify the exchange interactions by light over 1 %.

The symmetric in spin part of the exchange interaction is responsible for the very existence of magnetic ordering[5]. It is described by the Hamiltonian $\hat{H} = J \sum_{i,j} \left( \hat{\mathbf{S}}_i \cdot \hat{\mathbf{S}}_j \right)$, where $J$ is the exchange integral; $\hat{\mathbf{S}}_i$ and $\hat{\mathbf{S}}_j$ are the spins of the $i$th and $j$th adjacent magnetic ions. The antisymmetric part $\hat{H} = \mathbf{D} \cdot \sum_{i,j} \left( \hat{\mathbf{S}}_i \times \hat{\mathbf{S}}_j \right)$, called Dzyaloshinskii-Moriya interaction, gives rise to canted antiferromagnetism[6], magneto-electricity[7-8], chiral magnetic structures[9] and skyrmions[10-13].

The ability to control the exchange interaction by light has intrigued researchers in many areas of physics, ranging from quantum computing[14] to strongly correlated materials[15-17]. Laser-induced heating[18-19] and photo-doping[17,20] were suggested to cause a modification of the exchange interaction; however, these phenomena rely on absorption of light and are neither universal, i.e. only present in rather specific materials, nor direct, i.e. not instantaneous. The time-resolved evolution of the exchange splitting in magnetic metals Ni and Gd subjected to ultrafast laser excitation was measured using photoelectron spectroscopy[21] and angle-resolved photoemission techniques respectively. Both of these techniques, unfortunately, do not allow distinguishing the intrinsic dynamics of the exchange parameters such as $J$ from the demagnetization dynamics. Nevertheless, a direct truly ultrafast effect of the electric field of light on the exchange interaction must be feasible in any material. In a medium of arbitrary symmetry, such an effect may be expressed by introducing an isotropic term in the Hamiltonian of the two-photon interaction between the light and spin system

$$\hat{H} = I_{opt}\,\alpha \sum_{i,j}\left(\hat{\mathbf{S}}_i \cdot \hat{\mathbf{S}}_j\right) + 2\,I_{opt}\boldsymbol{\beta} \cdot \sum_{i,j}\left(\hat{\mathbf{S}}_i \times \hat{\mathbf{S}}_j\right),\tag{1}$$

where $I_{opt}$ is the intensity of light; $\alpha$ and $\boldsymbol{\beta}$ are some scalar and vector coefficients, respectively, which are defined by microscopic parameters. The presence of the interaction Hamiltonian (1) manifests itself as a magnetic refraction, described by an isotropic contribution to the dielectric permittivity $\varepsilon_{IMR} = aM^2$ that leads to a dependence of the refractive index on the magnetization $M$[23-24]. The first term in the Hamiltonian describes the intensity dependent contribution, $\Delta J = \alpha I_{opt}$, to the symmetric Heisenberg exchange integral $J$, whereas the second term describes the intensity dependent contribution $\Delta \mathbf{D} = \boldsymbol{\beta} I_{opt}$ to the Dzyaloshinskii-Moriya vector $\mathbf{D}$[6]. Recently the isotropic magneto-refraction effect has been utilized to probe the d-f exchange in EuTe[25]. As all other magneto-optical phenomena, the magneto-refraction must be connected with an inverse effect[26] described by the same Hamiltonian (1), i.e. the optical generation of a torque $\mathbf{T}_i$ acting on a spin $\mathbf{S}_i$ due to the light-induced perturbation of the exchange parameters

$$\mathbf{T}_i = -\gamma \left[ \mathbf{S}_i \times \frac{\partial \hat{H}}{\partial \mathbf{S}_i} \right] = -\gamma \Delta J \left[ \mathbf{S}_i \times \mathbf{S}_j \right] - \gamma \left[ \mathbf{S}_i \times \left[ \mathbf{S}_j \times \Delta \mathbf{D} \right] \right], \qquad (2)$$

where $\gamma$ is the absolute value of the gyromagnetic ratio.

In a broad class of transition metal oxides the magnetic order is governed by indirect exchange via ligand ions (super-exchange)[5], which is defined by virtual charge-transfer transitions of electrons between ligands and magnetic ions. Hence, one can anticipate the feasibility of a direct effect of the electric field of light on the exchange energy via virtual or real excitation of specific optical transitions that modify the hopping of the electrons between electronic orbitals centered at the transition metal ions and oxygen ligands, respectively (see Fig. 1). This plausible microscopic scenario results in the phenomenology of Eq. (1) at the macroscopic level.

Antiferromagnetic iron oxides possessing weak ferromagnetism, such as iron borate $FeBO_3$, rare-earth orthoferrites $R$FeO$_3$ and hematite α-$Fe_2O_3$, are excellent candidates for observing such ultrafast optical modification of the super-exchange interactions. In these compounds the $Fe^{3+}$ (S=5/2, L=0) ions form two magnetic sublattices, the spins of which are antiferromagnetically coupled due to the symmetric exchange interaction[27]. The presence of the Dzyaloshinskii-Moriya antisymmetric exchange interaction leads to a slight canting of the spins from the antiparallel orientation by an angle of ~0.5-1°. The value of the canting is defined by the ratio $D/J$ between the antisymmetric and symmetric exchange parameters. Thus one could expect that an ultrafast optical perturbation of the exchange parameters could also change their ratio $D/J$ and thereby trigger the so-called quasi-antiferromagnetic resonance mode by the torque (2) [see Fig. 2 (a)]. This mode corresponds to oscillations of the magnitude of the weak magnetic moment without a change of its orientation[28]. According to Eqs. (1) and (2), one expects that the ultrafast optical perturbation of the exchange parameters in these weak ferromagnets is an isotropic mechanism, i.e. it can excite the quasi-antiferromagnetic resonance independently from the light polarization and propagation direction. The excited oscillating magnetic dipole in turn leads to the generation of THz radiation which can be measured using THz emission spectroscopy[29], as has been demonstrated before in experiments with ferromagnetic metals[30-32] and antiferromagnetic insulators $NiO$[33-36] and $MnO$[37]. In the present context, observation of THz emission due to laser excitation of the quasi-antiferromagnetic spin resonance via an isotropic mechanism would indicate an ultrafast manipulation of the exchange interactions.

In our experiments we studied a $FeBO_3$ single crystal plate cut perpendicularly to the $z$-crystallographic axis. The sample was illuminated by ~100 fs laser pulses with their carrier frequency centered at 1.55 eV (800 nm). We performed time-resolved detection of the THz radiation emitted from the sample in the direction of the $z$-axis (see Fig. 2 (a)). The waveforms generated at different temperatures are shown in Fig. 2 (b). We observe that the optical excitation of the sample generates quasi-monochromatic emission at a frequency of ~0.45 THz [Fig. 2 (c)]], which corresponds to the frequency of the quasi-antiferromagnetic mode in $FeBO_3$[38]. The amplitude of the oscillations gradually decreases as the temperature approaches the Néel temperature $T_N$ ~350 K. To confirm that the observed effect is also present in other weak ferromagnets, similar experiments were performed on a $TmFeO_3$ single crystal plate cut perpendicularly to the $z$-crystallographic axis (see Fig. 2(d)). Figure 2 (e) demonstrates that, in the latter case, the sample emitted radiation at the frequency of ~0.8 THz, which is the frequency of the quasi-antiferromagnetic mode[28] in $TmFeO_3$.

In order to investigate the isotropic character of the excitation mechanism, we performed systematic measurements on the fluence and polarization dependence and found that the oscillation amplitudes depend linearly on the intensity of the pump (see Supplementary note 1) and are insensitive to the pump polarization (see Supplementary note 2). These observations are in perfect qualitative agreement with the anticipated features of an isotropic mechanism of optical modification of the exchange interaction described by Eq. (1). We have also observed similar polarization-insensitive ultrafast optical excitation of the quasi-antiferromagnetic mode in the $y$- and $x$-cut samples of ErFeO$_3$, the $x$-cut YFeO$_3$, and in hematite $\alpha$-Fe$_2$O$_3$ (see Supplementary note 3). Importantly, in all these compounds the effect was clearly seen even at room temperature. The phase of the observed oscillations changed over $\pi$ with the reversal of the magnetization direction, confirming the magnetic origin of the signals (see Supplementary note 4).

The consistent observation of the photo-excitation of the quasi-antiferromagnetic mode in a range of compounds clearly indicates that this effect originates from the perturbation of the $D/J$ ratio. Indeed, the observation of this effect in FeBO$_3$, which lacks significant in-plane anisotropy, rules out any substantial contribution from an optical modification of the magneto-crystalline anisotropy. At the same time, the isotropic and polarization insensitive character of the excitation rules out mechanisms based upon the inverse Faraday effect[1] which is sensitive to the ellipticity of the pump or the inverse Cotton-Mouton effect[39] which is sensitive to the polarization of the pump with respect to the magnetization direction. It is important to note here that the THz emission observed from antiferromagnets NiO[33-36] and MnO[37] did not contain a contribution isotropic with respect to the pump. Indeed the Dzyaloshinskii-Moriya antisymmetric exchange interaction is not allowed in these cubic insulators NiO and MnO and the torque (2) is equal to zero in accord with our model.

In order to specify the possible optical transitions responsible for our observations, we note that the excitation photon energy is not in resonance with any of the weak localized *d-d* crystal-field transitions in the $Fe^{3+}$ ions in $FeBO_3$ and orthoferrites[40-41]. However, the dispersion of the refraction coefficient for all these compounds is dominated by the off-resonant susceptibilities related to the electric-dipole allowed charge-transfer transitions between the $2p$ orbitals of oxygen and the $3d$ orbitals of the $Fe^{3+}$ ions above 3 eV[42-43]. During the laser-pulse duration and the time of optical decoherence, the collective electron wave-functions are coherent superpositions of the wave-functions of the ground and excited states. Such ultrafast modification of the wave-functions affects the exchange interaction between the spins of the neighbouring $Fe^{3+}$ ions and thus changes the energy of the super-exchange interaction (see Fig. 1). One can therefore expect that the observed effect of light on the exchange interaction is inherent to all magnetic materials, the magnetic order of which is governed by the super-exchange. However, only when the spins are canted either by the Dzyaloshinskii-Moriya interaction or by an applied magnetic field, such an ultrafast change of the exchange interaction will lead to excitation of the antiferromagnetic resonance and the subsequent emission of THz radiation in accord with Eq. (2). In materials with collinear magnetic configurations the torque (2) is zero since $\left[\mathbf{S}_i \times \mathbf{S}_j\right] = 0$.

Our data are in excellent agreement with the phenomenology of Eq. (1) that gives the simplest and most plausible explanation. A possible microscopic scenario underpinning the phenomenology of the obtained results can be understood in the framework of a recently developed formalism[44] for microscopic magnetic interactions out of equilibrium (see Methods sections and Supplementary note 6). To demonstrate the effect of a femtosecond laser pulse on the super-exchange interaction we numerically evaluated the time-dependent exchange for a 3-ion $Fe^{3+}$-$O^{2-}$-$Fe^{3+}$ cluster, which is characterized by a strong on-site Coulomb interaction $U$ on the $Fe^{3+}$ ions, an energy level shift $\Delta$ between the $Fe^{3+}$ and $O^{2-}$ ions, and an equilibrium hopping amplitude $t_0$ between Fe and O ions. For small ratio $t_0/U$, the leading-order expression for equilibrium super-exchange in this system[45] reads

$$J = \frac{2t_0^4}{U_1^2}\left(\frac{1}{U} + \frac{1}{U_1}\right),$$ where $U_1 = U + \Delta$. By gradually switching on an oscillating off-resonant

electric field we observe an enhancement of the exchange interaction proportional to the intensity of the laser pulse (see Supplementary note 6, Fig. S11). To further understand the dependence of the super-exchange on the laser field, we studied analytically a periodically driven cluster model. The shift of the energy levels under the periodic driving field can be understood within the Floquet theory (see Supplementary note 6), which gives an analytical expression for the change of the exchange interaction:

$$\Delta J = \frac{\varepsilon^2 t_0^4}{2}\left(\sum_{\pm}\left[\frac{1}{U_1 \pm \hbar\omega} + \frac{1}{U_1}\right]^2 \frac{1}{U \pm \hbar\omega} - \frac{4}{U_1^2 U} - \frac{4}{U_1^3}\right). \tag{3}$$

Here, $\varepsilon = eaE_0/\hbar\omega$ is the amplitude of the vector potential that describes the electric field in the Coulomb gauge with amplitude $E_0$, $e$ and $a$ are the unit charge and lattice constant, respectively, and $\omega$ is the frequency of the optical field. The terms dependent on $\pm\omega$ are the photon-assisted charge transfer excitations, while the last two terms describe a laser-induced decrease of the effective hopping amplitude within the $Fe^{3+}$-$O^{2}$-$Fe^{3+}$ cluster by a coherent destruction of tunneling[46]. We obtain excellent quantitative agreement of $\Delta J/J$ between Eq.(3) and the numerical results obtained from the general theory (see Supplementary note 6). In the experiment we typically have $\hbar\omega \sim U_1/2$, from which we conclude that the strengthening of the exchange interaction is caused by the photon-assisted charge-transfer excitation, as illustrated in Fig. 1. Using typical experimental parameters $U = 3$ eV, $\Delta = 0.25$ eV, $t_0 = 0.5$ eV and $\hbar\omega = 1.5$ eV, we find that an optical pulse with a fluence of 1 mJ/cm$^2$ and a corresponding electric field amplitude $E_0 = 0.12$ V/Å should induce a change of the exchange integral $\Delta J/J$ of over 1 %. Our model analysis neither incorporates multi-orbital effects nor a description of the non-equilibrium Dzyaloshinskii-Moriya interaction, which certainly would be beyond the scope of this report. However, the calculation of $\Delta J$ demonstrates in principle the plausibility of the proposed mechanism of optical manipulation of the symmetric exchange interaction. Importantly, we have shown theoretically that the optical manipulation of magnetic interactions is feasible already in the elementary super-exchange model defined by the Fe-O-Fe cluster. This suggests that the proposed microscopic mechanism is relevant in all magnetic systems governed by the super-exchange interaction, independently on the particular symmetry of their crystal structure.

To determine whether laser excitation leads to a decrease or an increase of the ratio $D/J$ we take advantage of the strong temperature dependence of the magnetic anisotropy, which is characteristic for many orthoferrites. For instance, heating of $TmFeO_3$ from 80 K to 90 K leads to a change of the equilibrium orientation of the weak magnetic moment from the $x$ to the $z$-axis. If the equilibrium orientation is changed as a result of a sudden heating by a femtosecond laser pulse, such a change is followed by oscillations of the weak magnetic moment in the $(xz)$-plane at the frequency of the quasi-ferromagnetic mode (~ 100 GHz)[47]. Our measurements clearly reveal that, in the range between 55 K and 68 K, such low-frequency oscillations corresponding to the quasi-ferromagnetic mode are observed in addition to the high-frequency quasi-antiferromagnetic oscillations (see Fig. 3). We applied a low pass filter to the data (cut-off frequency 250 GHz) to isolate the quasi-ferromagnetic mode and a high frequency filter (cut-off frequency 650 GHz) to isolate the quasi-antiferromagnetic mode. It is seen from Fig. 3 that the high-frequency mode measured at 60 K is in phase with that observed at 40 K. One can also see that the initial phases of the low-frequency quasi-ferromagnetic and high-frequency quasi-antiferrimagnetic modes are approximately $180^o$ apart. Note that for the $z$-cut $TmFeO_3$ sample, with a net magnetic moment oriented upwards, a laser-induced spin reorientation transition should trigger the quasi-ferromagnetic mode in such a way that the $M_x$-component of the magnetization decreases. The observed difference in the phases between the two oscillations shows that the quasi-antiferromagnetic mode is triggered in such a way that the $M_x$-component increases, meaning that the canting angle becomes larger. Such a behavior can only be explained by assuming that the quasi-antiferromagnetic oscillations are triggered by an *increase* of the ratio of the exchange parameters $D/J$. If this conclusion is true, in the *x*-cut sample the initial phases of the two modes must be the same, since the spin-reorientation in this sample goes in opposite direction. Measurements in the vicinity of the spin-reorientation temperature in ErFeO$_3$ cut perpendicular to the *x*-axis confirm this conclusion (see Supplementary note 5).

To deduce the magnitude and timescale of the exchange modification from the experimental data, we have solved the Maxwell equations for a slab of a material with an oscillating magnetization triggered by a perturbation of the ratio *D/J* and calculated the electromagnetic radiation emitted by the slab into the free space. A quantitative analysis supports the sub-picosecond impact on the spin system (see Supplementary notes 7 and 8). This implies that the light pulse changes the ratio $D/J$ directly via electric-dipole electronic transitions and not via the far slower heating via the lattice phonons. To further substantiate this conclusion we note that the specific heat and the thermal conductivity of the materials under consideration are strongly temperature dependent. For example, the specific heat of YFeO$_3$ below 100 K grows rapidly with increase of temperature while its thermal conductivity exhibits a pronounced peak around 30 K[48]. At the same time the efficiency of the quasi-antiferromagnetic mode excitation in this compound does not depend on temperature at all (see Fig. S4 (a) in Supplementary note 3) which indicates a minor role of heating in the excitation mechanism. The observation of the very same effect of comparable strength in hematite with high optical absorption ~ 2000 cm$^{-1}$ at 1.55 eV[49], in the orthoferrites with moderate optical absorption ~ 200 cm$^{-1}$ at 1.55 eV[40-41] and in virtually transparent iron borate with absorption ~ 50 cm[-1] at 1.55 eV[50] shows that the optical modification of the $D/J$ does not rely on optical absorption but originates from instantaneous light-matter interactions described by Hamiltonian (1).

The maximum value of the oscillating magnetization in the samples is estimated to be ~1 A/m. Oscillations with such an amplitude can only be triggered if the laser excitation results in an ultrafast increase of the ratio $D/J$ greater than 0.01 % (see Supplementary notes 7 and 8). Taking into account the parameters of our experiment, one can find that the sub-picosecond laser excitation with a fluence of ~ 1 mJ/cm$^2$ changes the potential energy of the magnetic system by ~ 1 μJ/cm$^2$ and acts as an effective magnetic field pulse of ~0.01 Tesla (see Supplementary note 9). These values (normalized to the optical fluence) correspond to some of the largest effects of light on magnetic systems observed to date[1,4].

To summarize, the demonstrated feasibility of sub-picosecond modification of the fundamental exchange parameters $J$ and $D$ and the ratio between them opens novel prospects for optical control of magnetically ordered materials. The suggested mechanism is not restricted by any requirement on the crystal symmetry and must thus be applicable to other classes of magnetic materials. Given that in some of them the isotropic magneto-refraction can be significantly larger than in iron oxides, we foresee many opportunities to enhance the effects reported here. We also anticipate that by tuning the wavelength of light, one should be able to affect selectively different exchange parameters in magnetic materials.

**Methods**

*Samples.* Single crystal plates with different thicknesses and crystallographic orientations were used in the measurements. The orthoferrite samples were 60-70 μm thick and cut perpendicularly to the *z*-axis (TmFeO$_3$), the *x*-axis (YFeO$_3$) and the *x*- and the *y*-axes (ErFeO$_3$). The iron borate FeBO$_3$ sample (370 μm) and hematite α-Fe$_2$O$_3$ sample (500 μm) were cut perpendicularly to the *z*-axis. The lateral size of all plates was ~ 5 mm.

*THz spectrometer.* A conventional time-domain THz spectrometer was used in the measurements. The THz spectrometer was powered by a Ti:sapphire amplified laser, emitting a sequence of optical pulses (800 nm wavelength (1.55 eV), 100 fs duration) with the repetition frequency of 1 kHz. Each laser pulse was divided into a stronger pump pulse and a weaker probe pulse. The pump spot size was larger than the aperture in the sample holder (~2 mm in diameter) to provide a quasi-uniform excitation with a fluence of ~1 mJ/cm$^2$. The electric field of the emitted THz wave was measured by the electro-optical sampling technique. The sample was held inside a closed cycle, helium cryostat (15-300 K, $10^{-4}$ mbar).

*Quantum theory of non-equilibrium exchange.* In this formalism quantum spin-spin interactions are obtained from a purely electronic model by introducing small time-dependent rotations of the spin-quantization axis. By integrating over the electronic degrees of freedom in the rotated reference frame, an effective quadratic spin model is obtained in which the spin-spin time-dependent interaction parameters are given in terms of non-equilibrium electronic Green functions and self-energies[39].

## Acknowledgements

The research leading to these results was partially funded from the European Commission's 7th Framework Program (FP7/2007-2013) under Grant Agreement 228673 (MAGNONICS) and from EPSRC of the UK under project EP/E055087/1. The work was partially supported by the Netherlands Organization for Scientific Research (NWO), the Foundation for Fundamental Research on Matter (FOM), the European Union's Seventh Framework Program (FP7/2007-2013) Grants No. NMP3-LA-2010-246102 (IFOX), No. 280555 (Go-Fast), No. 214810 (FANTOMAS), the European Research Council under the European Union's Seventh Framework Program (FP7/2007-2013)/ERC Grant Agreement No. 257280 (Femtomagnetism) and ERC Grant Agreement 339813(EXCHANGE). The funding support by the Russian Ministry of Education and Science was realized via the Program "Invited Scientist" (Grant Agreement 14.B37.21.0899) and the "Leading Scientist" Program (Project 14.B25.31.0025). R.V.P. and A.W. acknowledge the support of the joint RFBR-CAS project (No. 12-02-91172-a). A. W. thanks for the support of NSFC project (No. 50932003) and NSFC-NWO joint project (No. 51211130596). A. S. and M. I. K. acknowledge support from EU Seventh Framework Program grant agreement No. 281043 (FEMTOSPIN). J. H. M. acknowledges funding from NWO by a Rubicon grant.

## Author contributions

R.V.M., V.V.K. and A.V.K. conceived the project. E.H. and R.V.M. designed and built the experimental set up used to study $TmFeO_3$ and $YFeO_3$ samples at the University of Exeter. R.V.M. designed and built the experimental set up at the Radboud University Nijmegen to study the $ErFeO_3$, $FeBO_3$ and $\alpha$-$Fe_2O_3$ samples. R.V.M. performed all measurements, analyzed the data and developed the macroscopic theoretical formalism. A.S. and M.I.K. derived the quantum theory of non-equilibrium exchange. J.M. and M.E. derived the Floquet theory and performed numerical simulations of non-equilibrium exchange. R.V.P. and A.W. prepared the samples. R.V.M. and A.V.K co-wrote the paper with substantial contributions from E.H., V.V.K., R.V.P., A.S., J.M., M.E., M.I.K. and Th. R. The project was coordinated by A.V.K.

**Figures**

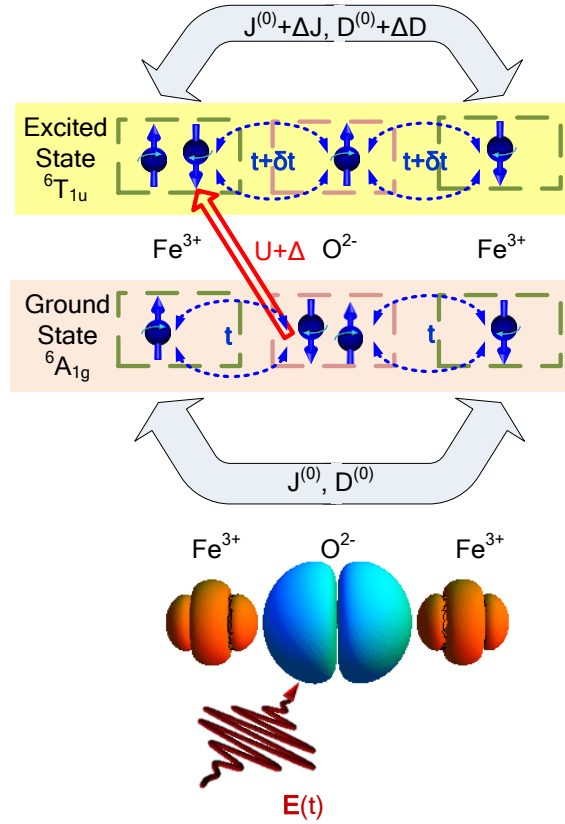

**Fig. 1. The mechanism of optical excitation of the exchange-driven spin dynamics in iron oxides.** The exchange interaction between the iron $Fe^{3+}$ ions (S=5/2, L=0) is mediated by the oxygen $O^{2-}$ ions and occurs due to the virtual hopping $t$ of electrons within the iron-oxygen complex. The laser pulse with photon energy 1.55 eV of arbitrary polarization excites charge-transfer virtual electric-dipole transitions $^6A_{1g}{\rightarrow}^6T_{1u}$ over the energy gap $U + \Delta$ in the iron-oxygen complex, thereby changing the electronic wave functions and modifying the ratio $D/J$.

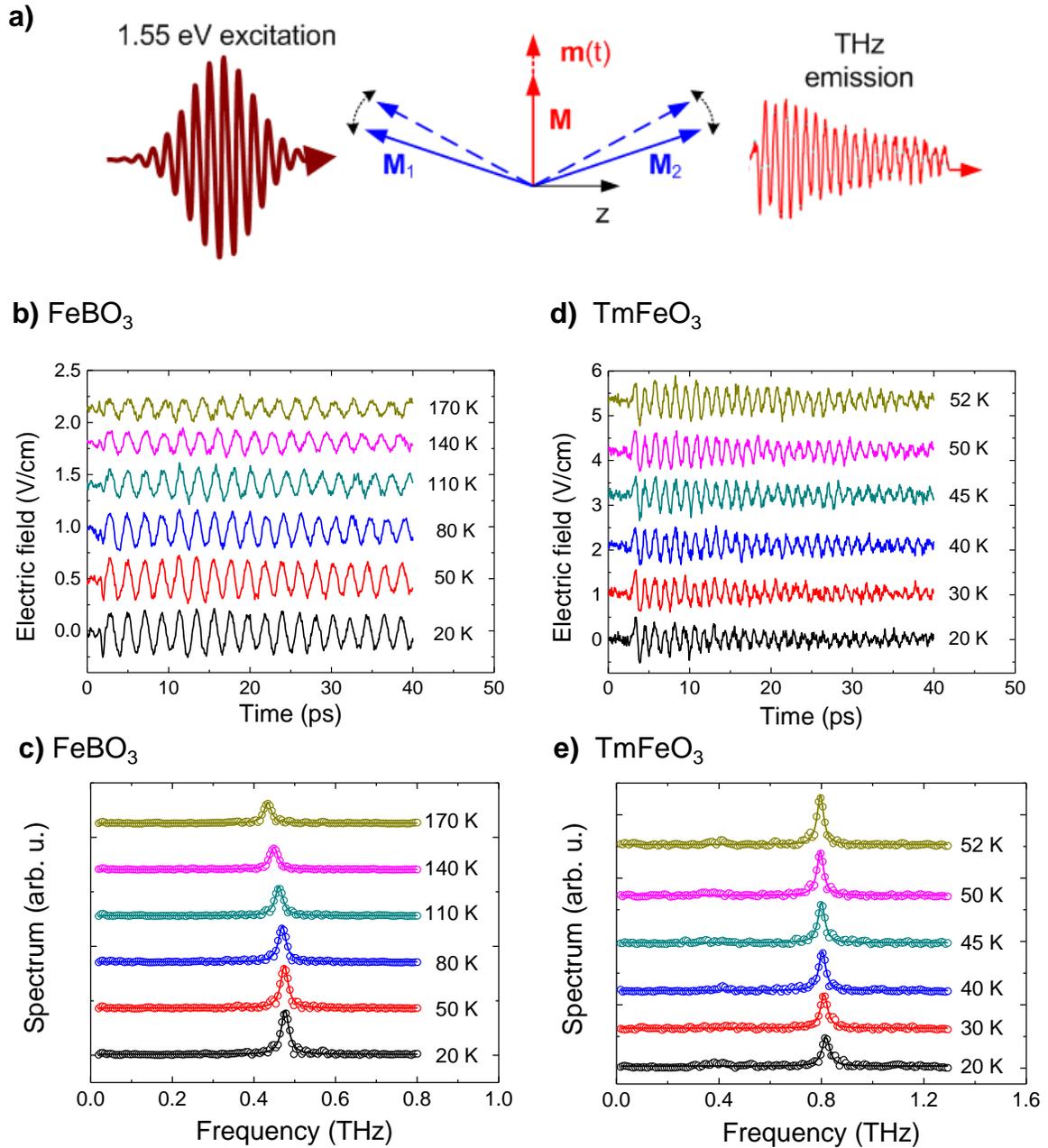

**Fig. 2. Terahertz emission generated in FeBO₃ and TmFeO₃ by 100 fs laser pulses. a**. The magnetization $\mathbf{M} = \mathbf{M_1} + \mathbf{M_2}$ lies in the plane of the crystal sample plate. The optical pump is focused onto the sample plate along its normal ($z$ axis), while the THz emission is collected along the same direction at the opposite side of the sample. The THz emission arises from the quasi-antiferromagnetic oscillations $\mathbf{m}(t)$. **b**. The FeBO₃ emission at different temperatures below 170 K. The zero time delay corresponds to the arbitrary starting position. The laser pulse arrives just before

the commencement of the oscillations. **c**. The spectra of the $FeBO_3$ emission obtained from the data by Fourier transform (open circles) fitted by Lorentzian functions (solid lines). **d**. The $TmFeO_3$ emission at different temperatures below 55 K. **e**. The $TmFeO_3$ emission spectra (open circles) fitted with Lorentzian functions (solid lines).

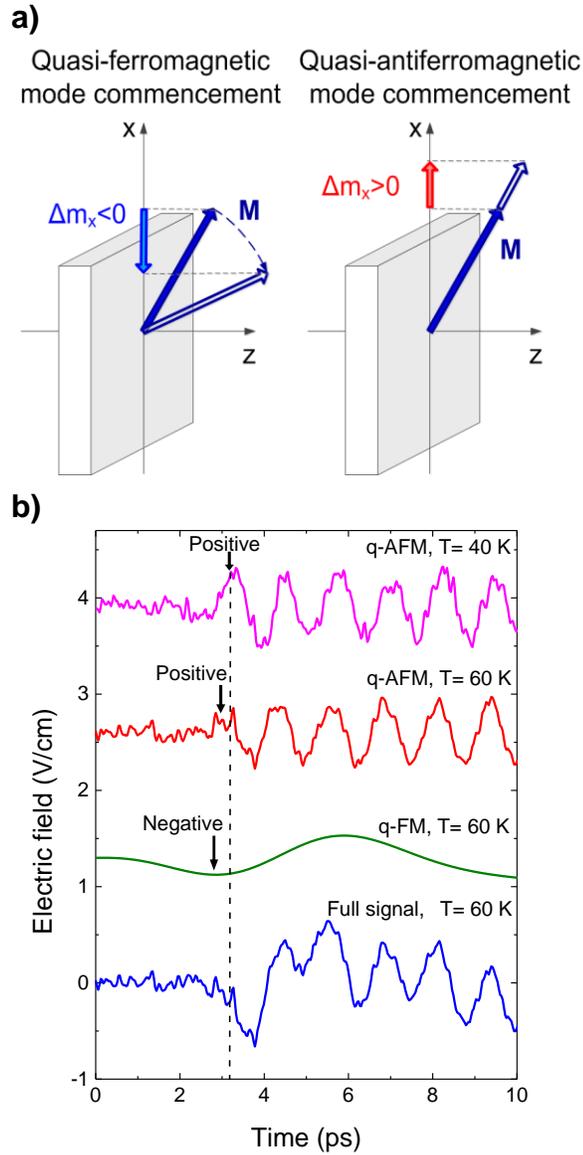

**Fig. 3. Determination of the absolute sign of the signal. a.** In the temperature interval 80-90 K, the spin configuration of TmFeO$_3$ continuously rotates in the ($xz$) plane, while keeping the weak ferromagnetic moment in the same plane. At the low temperature, the magnetization is oriented along the $x$-axis. So, due to the laser-induced reorientation, the $x$-component of the magnetization decreases initially (left panel). At the same time, due to the ultrafast change of the ratio $D/J$, the $x$-component of the magnetization increases (right panel). **b.** The signal emitted just below the spin-reorientation temperature at 60 K (green) is shown together with its low frequency part (green) and the high-frequency part (red). The latter part is in phase with the signal measured at 40 K which describes a quasi-antiferromagnetic mode only that commences out of phase with respect to the low frequency quasi-ferromagnetic oscillation (green curve).

# Supplementary notes

# Inverse magneto-refraction as a mechanism for laser modification of spin-spin exchange parameters and subsequent terahertz emission from iron oxides

R. V. Mikhaylovskiy, E. Hendry, A. Secchi, J. H. Mentink, M. Eckstein, A. Wu, R. V. Pisarev,

V. V. Kruglyak, M. I. Katsnelson, Th. Rasing and A. V. Kimel

**Supplementary note 1 – Signals generated in FeBO$_3$ and TmFeO$_3$ as functions of pump intensity and temperature**

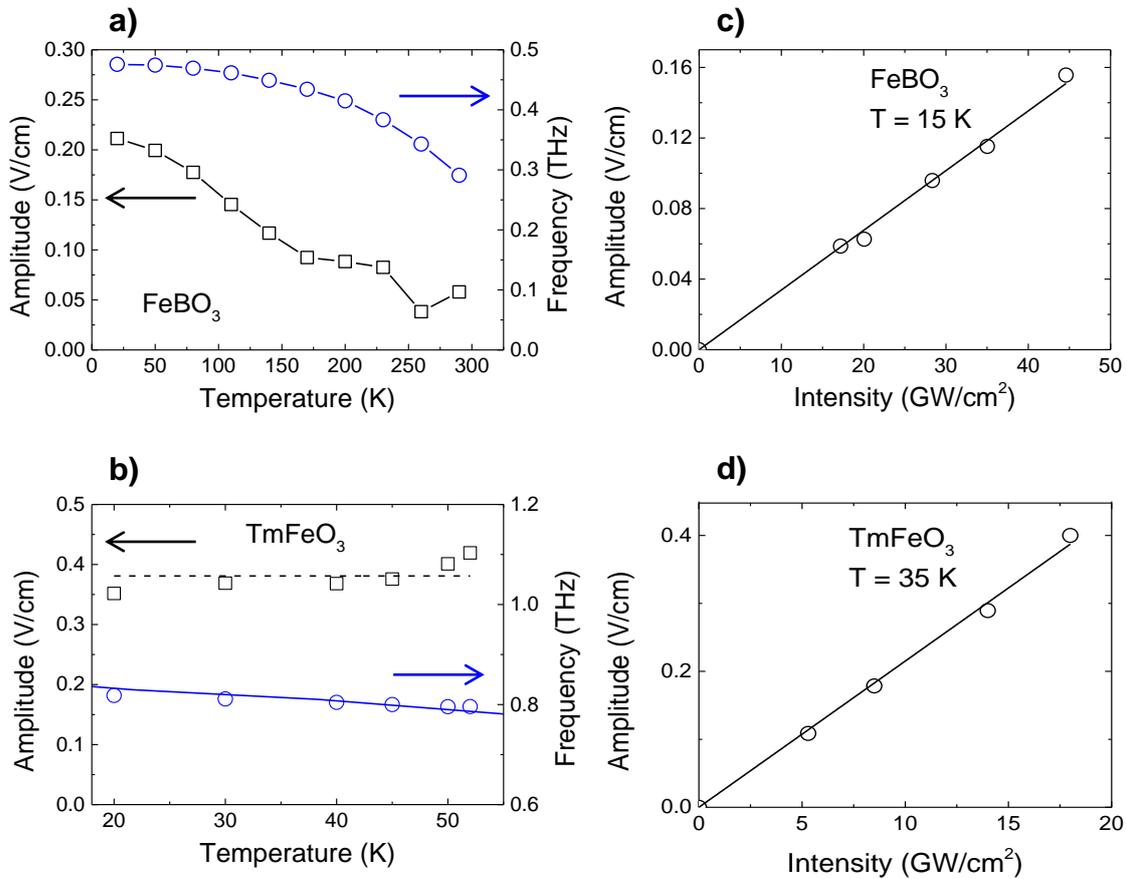

Fig. S1. Temperature and intensity behavior of the THz emission. The amplitude (open squares) and frequency (open circles) of the THz emission arising from the quasi-antiferromagnetic oscillation in FeBO$_3$ are shown as a function of temperature (a). The amplitude (open squares) and frequency (open circles) of the THz emission arising from the quasi-antiferromagnetic oscillation in TmFeO$_3$ vs temperature (b). The dashed line is a guide to the eye. The solid line shows the expected behavior of the quasi-antiferromagnetic resonance frequency extracted from Ref. 1. Note that the frequencies measured in our sample differ by ~ 10 % from those of Ref 1. The amplitude of the quasi-antiferromagnetic mode in FeBO$_3$ at 15 K (c) and in TmFeO$_3$ at 35 K (d) is a linear function of the pump intensity.

Fig. S1 (a) and (b) shows the amplitudes of the excited quasi-antiferromagnetic modes in FeBO$_3$ and TmFeO$_3$, respectively, as functions of temperature. The amplitude of the signal generated in TmFeO$_3$ is temperature independent, which agrees with the fact that the exchange interaction is weakly sensitive to temperature far from the Neel point ~ 650 K. At the same time the signal generated in $FeBO_3$ decreases as the temperature approaches the Neel point ~ 350 K in this compound. Such a decrease is expected provided that the magnetization of $FeBO_3$ becomes smaller at higher temperatures. The frequencies of the modes in both $TmFeO_3$ and $FeBO_3$ match the known frequencies of the quasi-antiferromagnetic resonance in these materials. The oscillation amplitudes are linearly dependent on the intensity of the pump [(Fig. S1 (c) and (d)] in accord with the Eq. (1) of the main paper.

**Supplementary note 2 – The THz emission generated in FeBO₃ and TmFeO₃ for different pump polarizations**

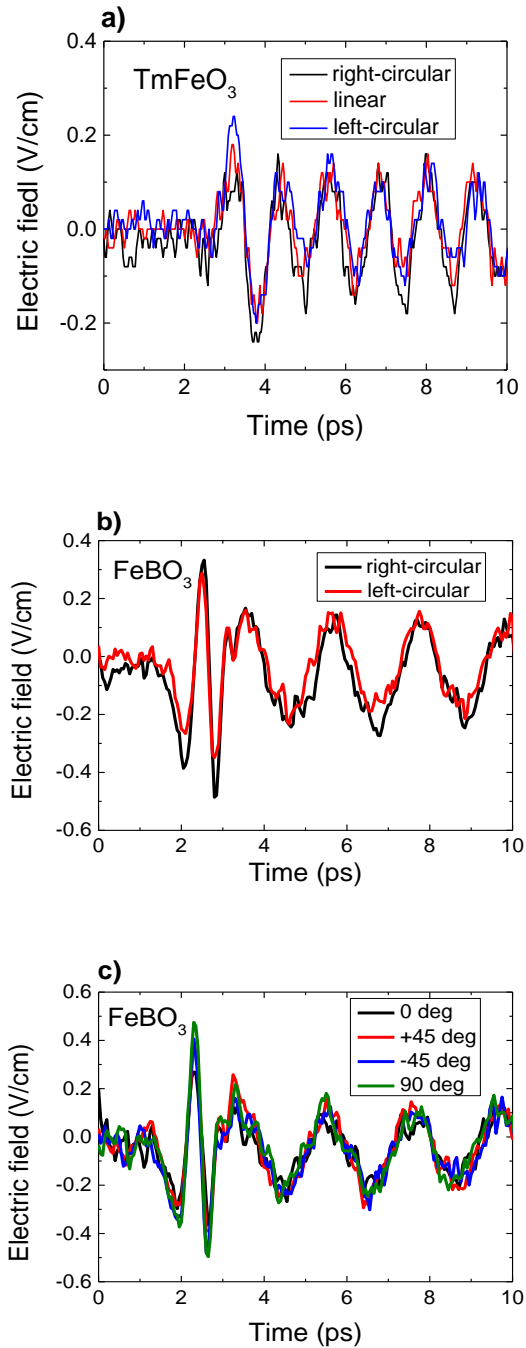

Fig. S2. The THz waveforms generated in TmFeO₃ (a) and FeBeO₃ (b), (c) by optical pulses with linear or circular polarizations. The angle in the legend of (c) corresponds to the orientation of the linear polarization with respect to the direction of the magnetization in the sample. The broadband pulse supposedly generated via electro-optical rectification arising from the surface nonlinearity is also seen at the start of the waveforms generated in FeBO₃. It indicates that the pump pulse arrival time equals ~ 1.5 ps

Figure S2 shows examples of THz emission associated with the quasi-antiferromagnetic mode in $TmFeO_3$ (a) and $FeBO_3$ (b), (c) for different pump polarizations. In the chosen geometries we did not find any signatures of polarization-sensitive effects within the whole temperature range available in the measurements.

**Supplementary note 3 – Additional data for emission from YFeO₃, ErFeO₃ and α-Fe₂O₃**

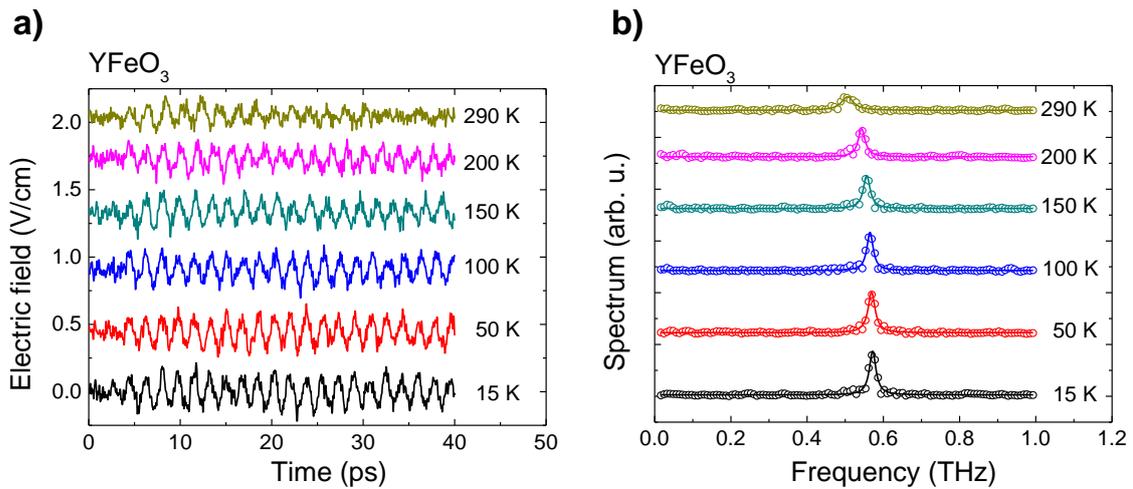

Fig. S3. The YFeO₃ emission waveforms (a) and the associated spectra (b) at different temperatures below room temperature.

The waveforms and their spectra generated in a YFeO₃ single crystal cut perpendicularly to the *x*-crystallographic-axis are shown in Fig. S3. The figure shows that, similarly to the case of FeBO₃ and TmFeO₃, using an ultrafast optical excitation we are able to excite oscillations at the frequency of ~0.55 THz, which again corresponds to the frequency of the quasi-antiferromagnetic mode in YFeO₃ [1].

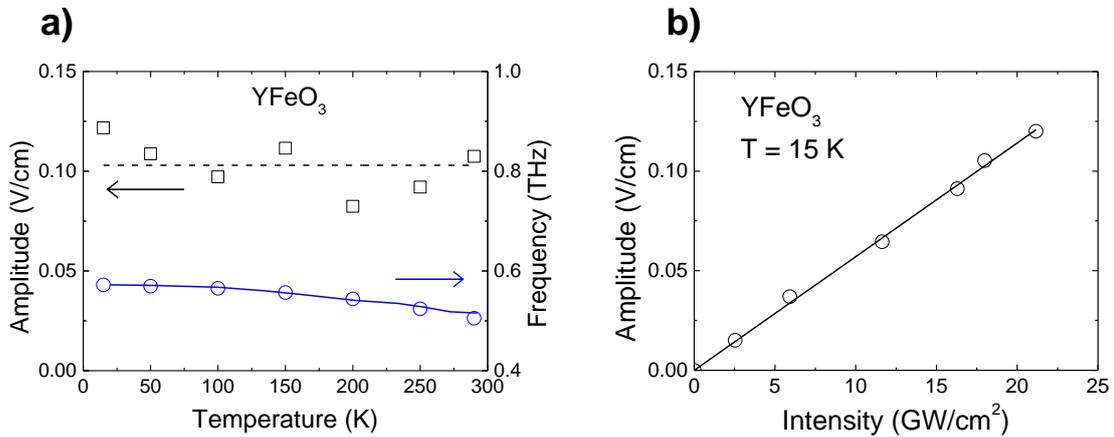

Fig. S4. The amplitude (open squares) and frequency (open circles) of the THz emission from YFeO₃ vs temperature is shown (a). The dashed line is a guide to the eye. The solid line shows the quasi-antiferromagnetic resonance frequency as a function of temperature taken from Ref. 2 with a correction shift of 60 GHz. The amplitude of the quasi-antiferromagnetic mode in YFeO₃ at 15 K is a linear function of the pump intensity (b).

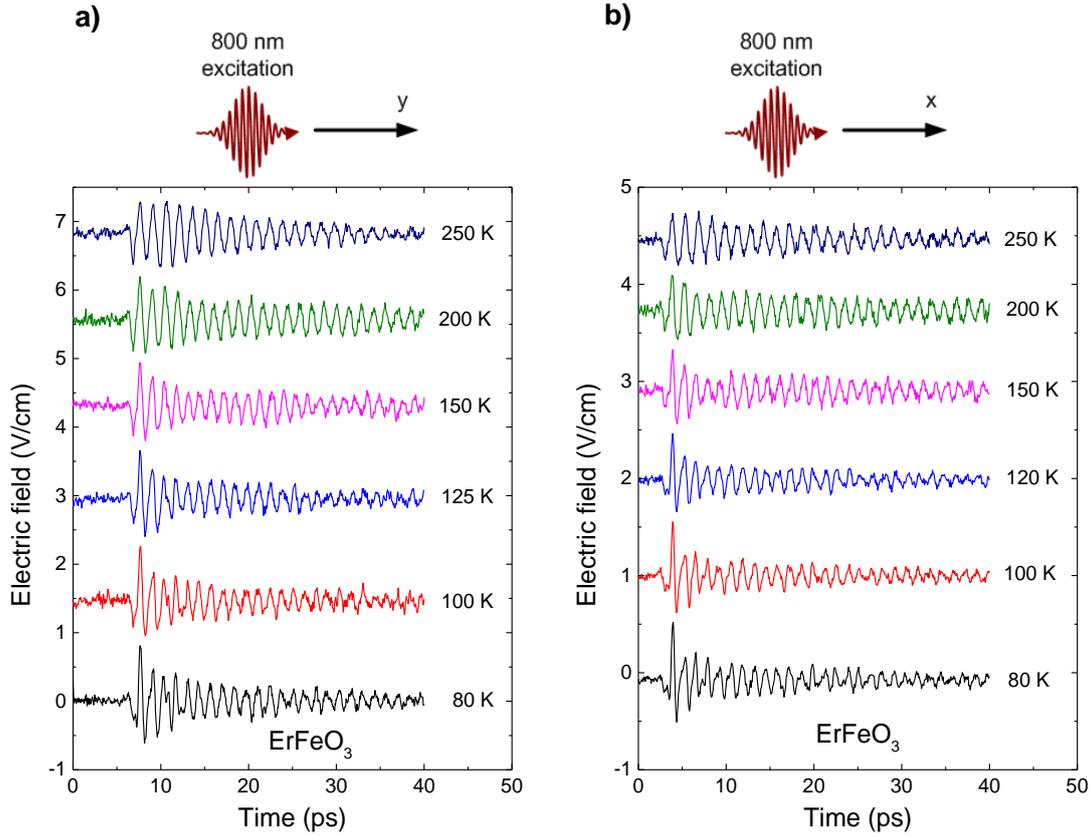

Fig. S5. The ErFeO$_3$ emission at different temperatures below room temperature generated by the optical pulses propagating along the $y$-axis (a) and the $x$-axis (b). At the temperatures below 150 K the high frequency rare-earth modes of Er interfere with quasi-antiferromagnetic mode of iron sub-lattices.

The amplitude of the emission signal generated in YFeO$_3$ is independent of temperature being a linear function of the light intensity (Fig. S4). The THz emission signals arising from quasi-antiferromagnetic oscillations (~0.8 THz [1]) in ErFeO$_3$ cut perpendicularly to the $y$ and $x$ axes (Figs. S5 (a) and (b) respectively) are very similar to each other thereby proving that the emission does not depend on the propagation direction of the pump pulse.

Fig. S6 (a) shows the examples of the THz electric field generated in α-Fe$_2$O$_3$. It is seen from the figure that the form of the signal is rather different from the quasimonochromatic signals generated in orthoferrites and iron borate. However, this is not surprising since the hematite crystal is highly absorbing at 800 nm, in contrast to the other materials measured in our experiments. The signal is dominated by the broadband THz pulse and its multiple Fabry-Perot replicas. This broadband pulse arises from either surface nonlinearity or magnetic dipole-electric quadrupole nonlinearity although investigation of its exact origin is beyond the scope of this study. However, one can observe [see Fig. S6 (b)] that above 250 K a spectral component at ~0.2

THz arises that is not present in the spectrum of the THz transient. The frequency of this component matches perfectly the frequency of the quasi-antiferromagnetic resonance in hematite [3]. Moreover, this spectral component does not depend on the pump polarization, while the rest of the emission spectrum does depend on the polarization of the pump which indicates their different origins. Importantly, the α-hematite exhibits a first-order phase transition from a purely antiferromagnetic state to the canted state at the critical temperature (Morin temperature) $T_M = 250$ K. Thus, only in the canted state above $T_M$ one anticipates the excitation of the quasi-antiferromagnetic resonance due to the modification of the exchange parameters. Indeed, the quasi-antiferromagnetic mode is observed only above the Morin point (Fig. S7).

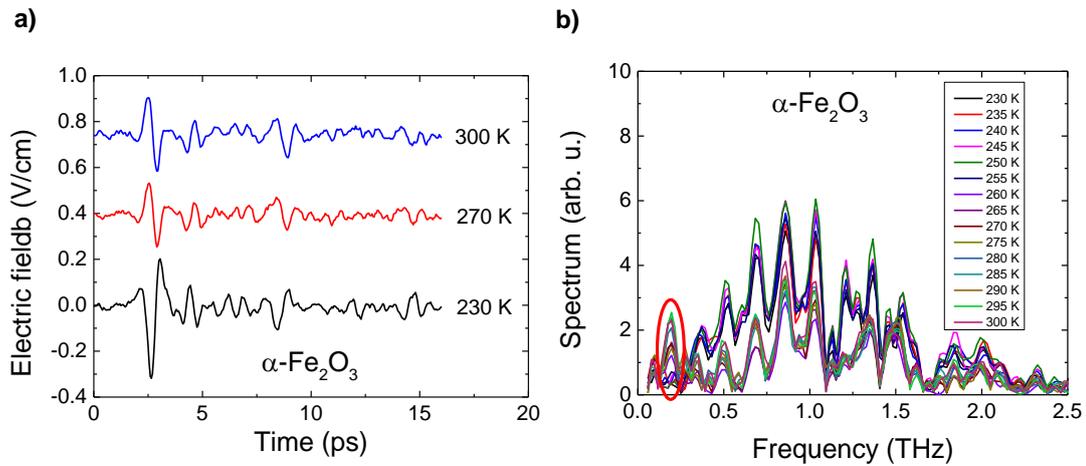

Fig. S6. The examples of the waveforms generated in α-hematite (a) and the spectra of the emission (b). The peak centered at ~ 0.2 THz arising only above the Morin point is marked with a red oval.

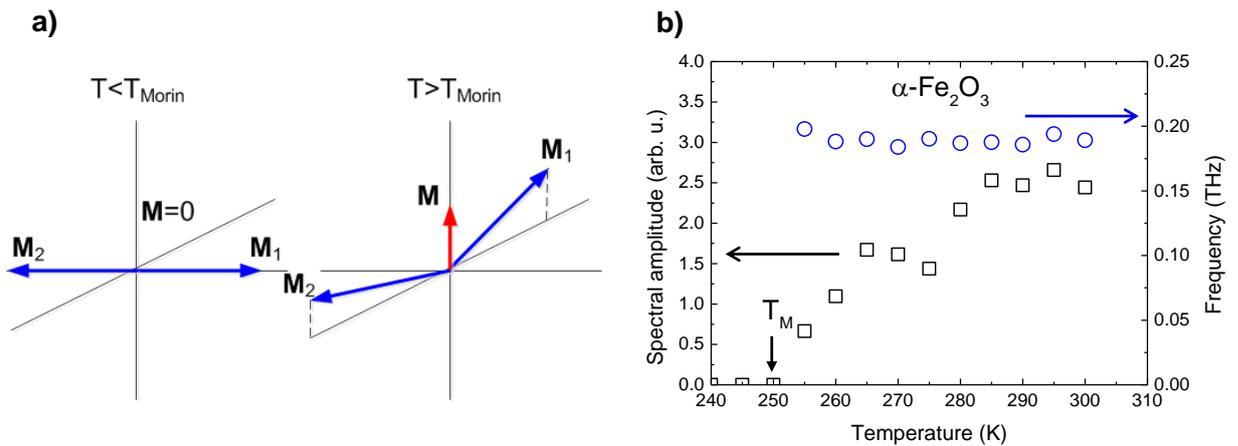

Fig.S7. The magnetic configuration of α-hematite below and above the Morin transition. (a). The amplitude (open squares) and frequency (open circles) of the THz emission from

α-Fe$_2$O$_3$ vs temperature (b). The quasi-ferromagnetic mode appears only in the canted antiferromagnetic phase above the Morin temperature $T_M$.

**Supplementary note 4 – The phase of THz emission is π-shifted upon inversion of the magnetization direction**

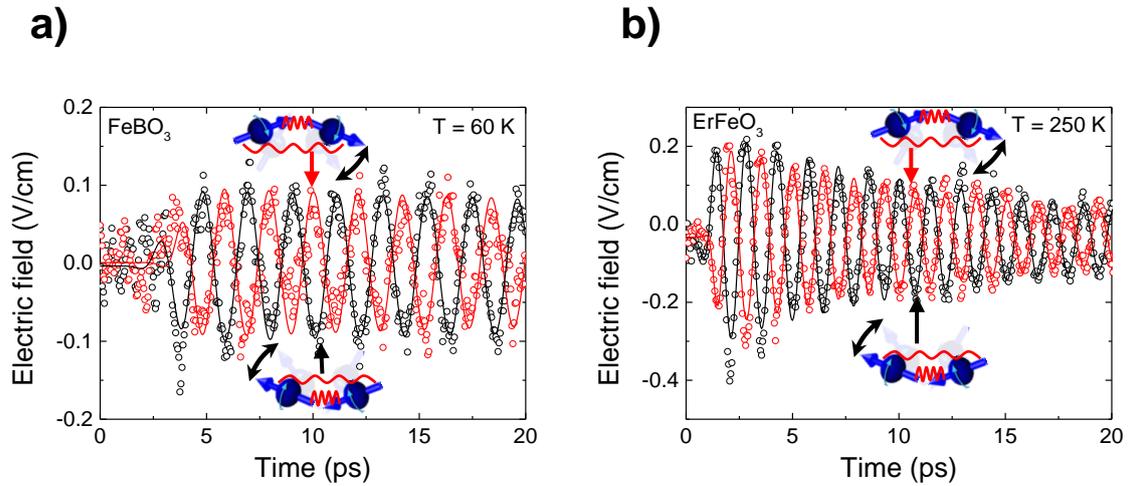

Fig. S8. The THz electric field emitted from FeBO$_3$ (a) and ErFeO$_3$ for opposite orientations of the magnetization (black and red open circles). The data are fitted with decaying sinusoids multiplied by error functions.

To further confirm the magnetic nature of the emitted radiation we checked that the phase of the observed oscillations is shifted over π as the magnetization reverses its polarity by the bias magnetic field ±1 kG (Fig. S8). Such a reversal of the sign of the signals proves that the THz oscillations arise from the spin precession.

**Supplementary note 5 – Does light increase or decrease $D/J$ in ErFeO$_3$?**

The magnetocrystalline anisotropy of TmFeO$_3$ and ErFeO$_3$ is characterized by a strong temperature dependence in the ~ 80 – 100 K temperature interval [4]. In this temperature range the spin configuration of the iron sub-lattices continuously rotates in the *(xz)* plane, while keeping the weak ferromagnetic moment in the same plane. Thus one might anticipate a strong temperature dependence of the THz emission in the vicinity of the spin-reorientation temperature interval. Indeed, along with the quasi-antiferromagnetic mode, another mode at ~ 100 GHz appears in the spectra of emission generated in TmFeO$_3$ and ErFeO$_3$ in vicinity of the spin-reorientation temperature range. We attribute this second mode to the quasi-ferromagnetic precession of spins.

The quasi-ferromagnetic resonance excitation under optical excitation of TmFeO$_3$ and ErFeO$_3$ near the spin reorientation temperature region has been reported before and assigned to the thermally induced change of the anisotropy [5-7]. This picture concurs with the fact that in YFeO$_3$ and FeBO$_3$ the low frequency mode has not been observed due to the absence of a spin reorientation in this material.

The analysis of the waveforms generated in TmFeO$_3$ at the temperatures of the photo-induced spin-reorientation allowed us to determine the absolute sign of the signal as discussed in the main text of the paper. It is instructive to check whether the sign of the quasi-antiferromagnetic oscillation generated in another compound exhibiting spin-reorientation, erbium orthoferrite, is consistent with the obtained result. We applied the same analysis to the signals generated in *x*-cut ErFeO$_3$ as illustrated in Fig. S9. Importantly, in such an oriented crystal the direction of the spin-reorientation is opposite with respect to the one in the *z*-cut crystalline plate of TmFeO$_3$. Thus, the initial phase of the quasi-antiferromagnetic mode must be the same as the initial phase of the low frequency quasi-ferromagnetic mode [Fig. S9 (a)]. This prediction has been fully validated by the experimental data, as shown in Fig. S9 (b).

## a)

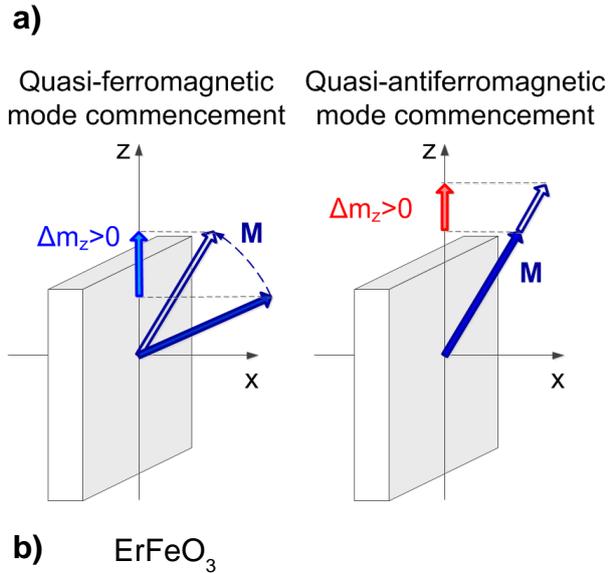

## b)

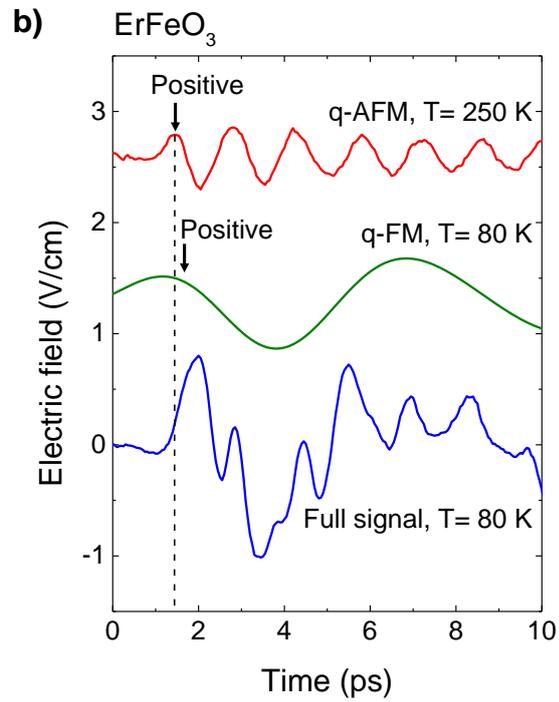

Fig. S9. Due to the laser-induced reorientation the z-component of the magnetization must increase initially (a – left panel). At the same time the z-component of the magnetization increases due to the ultrafast change of the ratio $D/J$ (a – right panel). The comparative analysis of the waveforms generated at the temperatures of the spin reorientation region and well above it demonstrates that the initial phases of the quasi-ferromagnetic and quasi-antiferromagnetic modes have the same sign (b).

**Supplementary note 6 – Microscopic theory of non-equilibrium exchange**

To demonstrate theoretically the feasibility of the modification of the super-exchange interaction, we adapt a quantum theory [8] that was recently developed to describe non-equilibrium magnetic interactions in strongly correlated systems. In particular, we specialize the application of this framework to a simple cluster model that mimics the experimental system (in particular, $\alpha$-Fe$_2$O$_3$) and we solve this model numerically. Furthermore, to provide additional theoretical understanding, we derive analytical results from Floquet theory for the same cluster model. Both the numerical and analytical results demonstrate an enhancement of the exchange interaction that scales linearly with the intensity of the electric field.

*Quantum theory of non-equilibrium magnetic interactions*

Our theory exploits the non-equilibrium Green-function formalism developed by Schwinger [9], Keldysh [10], Kadanoff and Baym [11]. The electronic partition function $Z$ is written as a path integral over fermionic (Grassmann) fields of the exponential of a non-equilibrium action $S[\bar{\phi}, \phi]$, i.e.,

$$Z = \int \mathrm{D}[\bar{\phi}, \phi] \exp\left(\mathrm{i}\, S[\bar{\phi}, \phi]\right). \tag{1}$$

The effective action describes the system in equilibrium for $t < 0$, going out of equilibrium for $t > 0$. Fermion fields $\phi_{a\sigma}$ are labelled by indices $a$ and $\sigma = \{\uparrow, \downarrow\} = \{+, -\}$, referring respectively to site (we are considering the single band case below) and spin single-particle states. The spin quantization axis of the $\phi$ fermions is along the unit vector $\mathbf{u}_z$. To study the spin excitations on top of the equilibrium ground state, we apply time- and site- dependent rotations to the local spinors, defined as $\phi_a = (\phi_{a\uparrow}, \phi_{a\downarrow})^T$, in order to map the old fermion fields $\phi_{a\sigma}$ into new fields $\psi_{a\sigma}$ having their spins aligned with time-dependent unit vectors $\sigma \mathbf{e}_a(t)$, which are interpreted as the directions of (classical) magnetic moments. The deviations of the $\mathbf{e}_a(t)$'s from the equilibrium direction $\pm \mathbf{u}_z$ are described by auxiliary (Holstein-Primakoff) boson fields; for low-energy excitations we assume such deviations to be small, obtaining an action quadratic in the bosons (this corresponds to small mixing between quantum states with different total spin). By integrating out the fermion fields $\psi$ an effective bosonic action is obtained and we finally map the bosons to the $\mathbf{e}_a(t)$ fields to explicitly identify the spin-spin interactions. The coefficients describing the magnetic interactions are expressed in terms of non-equilibrium electronic Green's functions and related self-energies, which generalize the equilibrium formalism [12] to include the effects of an external time-dependent field.

Assuming that the spin dynamics is slow with respect to electronic hopping processes, we obtain the following formula for the non-equilibrium exchange coupling between sites $a$ and $b$:

$$\eta_{ab}\, j_{ab}(t) = \frac{1}{4}\mathrm{Re}\left[F_{ab}(t)\right] + \frac{1}{4}\int_0^t dt'\, \mathrm{Im}\left[A_{ab}^>(t',t) - A_{ab}^<(t',t)\right], \qquad (2)$$

where $\eta_{ab} = +1\,(-1)$ if the ground-state spin correlation function of sites $a$ and $b$ is antiferromagnetic (ferromagnetic), and the quantities to be computed are the following (see Eqs. (107) and (124) in [8]):

$$A_{ab}^>(t,t') = \sum_c \sum_d \left[ -\delta_{ac}\, \mathrm{i}\frac{\overrightarrow{\partial}}{\partial t} + \left(1 - \overrightarrow{P}_{ac}\right) t_{ac}(t) \right] G_{cb}^{<\downarrow}(t',t)\, G_{da}^{>\uparrow}(t',t)$$
$$\times \left[ -\delta_{bd}\, \mathrm{i}\frac{\overleftarrow{\partial}}{\partial t'} + t_{bd}(t')\left(1 - \overleftarrow{P}_{bd}\right) \right]. \qquad (3)$$

$A_{ab}^<(t,t')$ is obtained from this equation by exchanging $>$ and $<$ in the right-hand side, and for $a \neq b$,

$$F_{ab}(t)\big|_{a \neq b} = \mathrm{i}\,\overline{\Sigma}_{ab}^\downarrow(t)\, G_{ba}^{<\uparrow}(t,t) + \mathrm{i}\, G_{ab}^{<\downarrow}(t,t)\,\overline{\Sigma}_{ba}^\uparrow(t). \qquad (4)$$

The above quantities are expressed in terms of the non-equilibrium Green's functions,

$$G_{ba}^{<\sigma}(t,t') = \mathrm{i}\left\langle \hat{\psi}_a^{+\sigma}(t')\, \hat{\psi}_b^\sigma(t) \right\rangle, \quad G_{ba}^{>\sigma}(t,t') = -\mathrm{i}\left\langle \hat{\psi}_b^\sigma(t)\, \hat{\psi}_a^{+\sigma}(t') \right\rangle, \qquad (5)$$

and in terms of the time-dependent Hartree-Fock component of the self-energy $\overline{\Sigma}(t)$. Furthermore, in Eq. (3) the arrows above the operators define the directions along which the operators act, $P_{ab}$ is the operator interchanging indices $a$ and $b$ in the functions it acts upon, and $t_{ab}(t)$ are hopping matrix elements, which depend on time in the presence of a time-dependent external field.

For the numerical calculations, it is convenient to work out the derivatives analytically, which yields the lesser and greater components of [1]

$$A_{ab}(t,t') = R_{ab}^{\downarrow}(t,t')\, R_{ba}^{\uparrow}(t',t) + S_{ab}^{\downarrow}(t,t')\, S_{ba}^{\uparrow}(t',t) - G_{ab}^{\downarrow}(t,t')\, T_{ba}^{\uparrow}(t',t)$$
$$- T_{ab}^{\downarrow}(t,t')\, G_{ba}^{\uparrow}(t',t).$$

(6)

Here we have introduced the quantities

$$R_{ab}^{\sigma}(t,t') = \left[\Sigma * G\right]_{ab}^{\sigma}(t,t'),$$

(7a)

$$S_{ab}^{\sigma}(t,t') = \left[G * \Sigma\right]_{ab}^{\sigma}(t,t'),$$

(7b)

$$T_{ab}^{\sigma}(t,t') = \Sigma_{ab}^{\sigma}(t,t') + \left[\Sigma * G * \Sigma\right]_{ab}^{\sigma}(t,t'),$$

(7c)

where $\Sigma_{ab}^{\sigma}(t,t')$ is the self-energy, which accounts for the electron correlations.

It must be noticed that the effective exchange interaction in Eq. (2) is not the bare exchange interaction between magnetic moments in sites $a$ and $b$, since it also includes a term which describes the variation with time of the magnitude of the local magnetic moments (non-Heisenberg effects). It can be shown that, in the absence of symmetry breaking (which is the case we will consider here), the bare exchange parameters are obtained as

$$J_{ab}(t) = \frac{j_{ab}(t)}{\left\langle \mathbf{s}_a(t) \cdot \mathbf{s}_b(t) \right\rangle},$$

(8)

where $\left\langle \mathbf{s}_a(t) \cdot \mathbf{s}_b(t) \right\rangle$ is the equal time spin-spin correlation function.

---

[1] The symbols $A^{>}$ and $A^{<}$ in Eq.(3) are defined according to the conventions used in Ref.[9]; in particular, time variables $t$ and $t'$ are defined on the real axis. In Eq.(6) time variables are defined on the Keldysh contour. The correspondence between Eqs.(3) and (6) is given by $A^{>}(t,t') = A(t_+,t'_-)$ and $A^{<}(t,t') = A(t_-,t'_+)$, where $\pm$ indicate times on the upper and lower branches of the Keldysh contour, respectively. Note that this is different from the conventional definition of lesser and greater components of Keldysh functions.

*Minimal model for super exchange*

As minimal model for super-exchange we consider a chain of three atoms denoted 0, 1 and 2 [13, 14]. Atoms 0 and 2 correspond to transition metal sites with one partially filled *d*-orbital and atom 1 contributes one filled (oxygen) *p*-orbital. The Hamiltonian consists of a local part $H_{\mathrm{loc}}$ and a time-dependent hopping term $H'(t)$:

$$H_{\mathrm{loc}} = \varepsilon_d \sum_\sigma (n_{0\sigma} + n_{2\sigma}) + \varepsilon_p \sum_\sigma n_{1\sigma} + U \sum_{j=0,2} n_{j\uparrow} n_{j\downarrow}, \tag{9a}$$

$$H'(t) = -\sum_{j=0,1} \sum_\sigma t_0 \, \mathrm{e}^{\mathrm{i}\varphi(t)} c^+_{j\sigma} c_{j+1\,\sigma} + h.c. \quad . \tag{9b}$$

Here $c^+_{j\sigma}$ creates an electron with spin $\sigma = \{\uparrow, \downarrow\}$ at site *j,* and $n_{j\sigma} = c^+_{j\sigma} c_{j\sigma}$ is the number operator. We choose the zero of the energy as $\varepsilon_d + \varepsilon_p = 0$, and define $\Delta = \varepsilon_d - \varepsilon_p$; $U$ is the local (Hubbard) interaction energy associated with *d* orbitals, that is, $U = \iint d\vec{r}\, d\vec{r}' \left| \psi_{j,d}(\vec{r}) \right|^2 V(\vec{r} - \vec{r}') \left| \psi_{j,d}(\vec{r}') \right|^2$, where $\psi_{j,d}(\vec{r})$ is the wave function of the *d* orbital localized on site *j* and $V(\vec{r} - \vec{r}')$ is the effective Coulomb interaction energy between electrons at positions $\vec{r}$ and $\vec{r}'$. The single-electron Hamiltonian $H'(t)$ accounts for time-dependent hoppings between *p* and *d* orbitals. Specifically, the equilibrium hopping parameter $t_0 = -\int d\vec{r}\, \psi^*_{0,d}(\vec{r})\, H_1(\vec{r})\, \psi_{1,p}(\vec{r})$, with the dimensions of energy, is the matrix element between a *d* and a *p* orbital (respectively localized on iron and on oxygen) of the equilibrium single-electron Hamiltonian $H_1(\vec{r})$, which includes the kinetic energy of the electrons, the interaction with the ions, as well as any other single-electron time-independent potential.

The time-dependent electric field $E(t)$ is included in $H'(t)$ (see Eq. 9b) by means of the time-dependent Peierls substitution [15- 17], which is equivalent to multiplying the equilibrium hopping by a time-dependent phase factor. In the Coulomb gauge (zero scalar potential) and for a spatially uniform vector potential the Peierls phase becomes

$$\varphi(t) = \frac{ea}{\hbar c} A_\parallel(t), \tag{10}$$

where $A_\parallel(t)$ is the component of the vector potential parallel to the chain and $a$ is the lattice spacing. The electric field is then related to the vector potential as $\mathbf{E}(t) = -\dfrac{1}{c}\dfrac{\partial}{\partial t}\mathbf{A}(t)$. For periodic driving field $E_\parallel(t) = E_0 \cos(\omega t)$ along the chain we obtain

$$\varphi(t) = -\varepsilon \sin(\omega t), \tag{11}$$

where the amplitude is given in terms of the dimensionless parameter

$$\varepsilon = \frac{eaE_0}{\hbar\omega}. \tag{11}$$

For the numerical solution of the periodically driven cluster model we slowly switch on the Peierls phase using an error function envelope

$$\varphi(t) = -\frac{1}{2}\varepsilon \sin(\omega t)\left[\mathrm{erf}(\alpha(t - t_1)) + 1\right]. \tag{12}$$

where (for a given $\omega$) the parameters $\alpha$ and $t_1$ are chosen such that $\varphi(0) = 0$ and the rise time takes about 10 oscillation periods.

*Numerical computation of non-equilibrium exchange parameters*

A numerically exact solution of the 3-site super-exchange model out of equilibrium is obtained by solving the time-dependent Schrödinger equation using exact diagonalization. From the time evolution of the states we evaluate the following correlation functions using the Lehmann representation:

$$G_{ab}^\sigma(t,t') = -\mathrm{i}\left\langle \hat{T}_C\, c_{a\sigma}(t)\, c_{b\sigma}^+(t') \right\rangle, \tag{13a}$$

$$R_{ab}^\sigma(t,t') = -\mathrm{i}\left\langle \hat{T}_C\, n_{a\bar{\sigma}}(t)\, c_{a\sigma}(t)\, c_{b\sigma}^+(t') \right\rangle U_a, \tag{13b}$$

$$S_{ab}^\sigma(t,t') = -\mathrm{i}\left\langle \hat{T}_C\, c_{a\sigma}(t)\, c_{b\sigma}^+(t')\, n_{b\bar{\sigma}}(t') \right\rangle U_a, \tag{13c}$$

$$T_{ab}^\sigma(t,t') = -\mathrm{i}\left\langle \hat{T}_C\, n_{a\bar{\sigma}}(t)\, c_{a\sigma}(t)\, c_{b\sigma}^+(t')\, n_{b\bar{\sigma}}(t') \right\rangle U_a\, U_b, \tag{13d}$$

where $\hat{T}_C$ is the time-ordering operator along the Keldysh contour and $U_a$ is the Coulomb interaction at site $a$. By computing products like $R_{ab}^{\downarrow}(t,t')R_{ba}^{\uparrow}(t,t')$ and by combining all terms we obtain the quantity $A_{ab}(t,t')$.

In addition, we evaluate the spin-spin correlation function as

$$\langle \mathbf{s}_a(t) \cdot \mathbf{s}_b(t) \rangle = \sum_\sigma \left[ \frac{1}{2} \langle c_{a\sigma}^+(t) \, c_{a\bar{\sigma}}(t) c_{b\sigma}^+(t) \, c_{b\sigma}(t) \rangle + \frac{1}{4} \langle n_{a\sigma}(t) \, n_{b\sigma}(t) - n_{a\sigma}(t) \, n_{b\bar{\sigma}}(t) \rangle \right]. \quad (14)$$

For the cluster model we consider a filling with 4 electrons and total spin $S_z = 0$, and prepare the system at $t = 0$ in the ground state at low temperature.

In equilibrium the exchange interactions are extracted at zero electric field by evolving the system along the real time axis to extract $A_{ab}(t,t') = A_{ab}(t - t')$ for $a, b = 0, 2$. The full (static) exchange interaction $j_{ab}^0$ is obtained from Eq.(2), by taking $t = t_*$ large enough,

$$j_{ab}^0 = j_{ab}(t_*). \quad (15)$$

We use a Gaussian window of length $L$ to ensure a smooth cutoff of the upper integration boundary. Numerically converged results, independent of $L$ and $t_*$, are obtained for $t_* > L > j_{ab}^{-1}$. The bare exchange interaction, Eq. (8) is then computed as

$$J_{ab}^0 = \frac{j_{ab}^0}{\langle \mathbf{s}_a(0) \cdot \mathbf{s}_b(0) \rangle}. \quad (16)$$

Out of equilibrium, the hopping matrix elements are modulated by the Peierls phase, and the $j_{ab}(t)$ and $\langle \mathbf{s}_a(t) \cdot \mathbf{s}_b(t) \rangle$ oscillate with the electric field. In this case, we compute the effective exchange interaction by averaging the bare exchange interaction over the period $T$ of the field

$$J_{ab}^\varepsilon = \frac{1}{T} \int_{t_*}^{t_*+T} dt \left[ \frac{j_{ab}(t)}{\langle \mathbf{s}_a(t) \cdot \mathbf{s}_b(t) \rangle} \right]. \quad (17)$$

The result is found to be independent on $t_*$ in the quasi-equilibrium state formed after slowly ramping up the Peierls phase.

*Analytical Floquet theory*

To gain additional understanding on the electrical control of the exchange interaction we use an alternative approach based on Floquet theory. For small amplitudes $\varepsilon$, relevant to the experimental conditions, this allows us to derive analytical formulae for the non-equilibrium exchange interaction under periodic driving of the 3-site cluster model.

To discuss the superexchange mechanism in the unperturbed system ($\varepsilon = 0$), we consider a filling of the cluster with 4 electrons, and apply perturbation theory to describe the low-energy states. For large Coulomb interaction ($U, U + \Delta >> t_0$) the low-energy sector of the Hilbert space includes the states with two electrons occupying the oxygen $p$-orbital and one electron in each of the two transition metal $d$-orbitals $H_0 = \left\{ \left| \downarrow, 2, \uparrow \right\rangle, \left| \uparrow, 2, \downarrow \right\rangle, \left| \downarrow, 2, \downarrow \right\rangle, \left| \uparrow, 2, \uparrow \right\rangle \right\}$. The symbols in $\left| \quad \right\rangle$ denote the occupation of the orbitals 0, 1, 2, from left to right, which can be either empty (0), singly occupied $(\uparrow, \downarrow)$, or doubly occupied (2). Virtual charge-transfer excitations due to the hopping $t_0$ to states $\left| 2, 2, 0 \right\rangle, \left| 0, 2, 2 \right\rangle$ etc. lead to a shift of the levels. A splitting $E_T - E_S$ of singlet $\left( \left| \downarrow, 2, \uparrow \right\rangle - \left| \uparrow, 2, \downarrow \right\rangle \right)/\sqrt{2}$ and triplet levels occurs in fourth order,

$$E_T - E_S = 4t_0^4 \left( \frac{1}{U_1^2 U} - \frac{1}{U_1^3} \right). \tag{18}$$

where $U_1 = U + \Delta$. The low-energy Hamiltonian is thus an antiferromagnetic Heisenberg model with exchange interaction $J = \left( E_T - E_S \right)/2$.

For a periodically driven system, solutions of the time-dependent Schrödinger equation can be found in terms of the Floquet modes of the form

$$\left| \psi(t) \right\rangle = e^{-i\zeta_\alpha t} \left| \psi_\alpha(t) \right\rangle, \tag{19}$$

where $\left| \psi_\alpha(t) \right\rangle = \left| \psi_\alpha(t + T) \right\rangle$ is periodic with $T = 2\pi/\omega$ [18-19]. To determine $\left| \psi_\alpha(t) \right\rangle$ and the Floquet spectrum $\zeta_\alpha$, we expand $\left| \psi_\alpha(t) \right\rangle$ as

$$\left| \psi(t) \right\rangle = \sum_n e^{-i\omega n t} \left| \psi_{\alpha,n}(t) \right\rangle. \tag{20}$$

The Schrödinger equation $i\hbar \, \partial_t \left| \psi(t) \right\rangle = H(t) \left| \psi(t) \right\rangle$ gives

$$(H_{\text{loc}} + n\hbar\omega)\big|\psi_{\alpha,n}(t)\big\rangle + \sum_m H'_m\big|\psi_{\alpha,n+m}(t)\big\rangle = \zeta_\alpha\big|\psi_{\alpha,n}(t)\big\rangle, \qquad (21)$$

where $H'_n$ are the Fourier components of the hopping term $H'(t)$,

$$H'_n = \frac{1}{T}\int_0^T dt\, H'(t)\,e^{i\omega nt}\ . \qquad (22)$$

The quasi-energies parametrically evolve with $\varepsilon$ from the unperturbed energies $\zeta_\alpha^0 = E_\alpha(+n\hbar\omega)$ to the perturbed levels $\zeta_\alpha(\varepsilon)$. As a function of time, the amplitude evolves with the pulse envelope, and thus has typically only small variation on the timescale of the period $T$. Under these conditions, the system would adiabatically evolve from the low energy states of $H(\varepsilon = 0)$ into the corresponding Floquet modes, and the slow dynamics during the pulse (in particular, precessional motion of spins) is governed by an effective Hamiltonian that is determined by the level spectrum $\zeta_\alpha(\varepsilon)$. In particular, we can obtain the non-equilibrium exchange interaction from the singlet-triplet splitting as $J(\varepsilon) = (\zeta_{\text{T}}(\varepsilon) - \zeta_{\text{S}}(\varepsilon))/2$.

The experiment is performed in the regime of weak perturbation $\varepsilon \ll 1$ (from the estimates in the main text we get $\varepsilon \sim 0.04$). In this case, the level splitting can be determined analytically by standard degenerate perturbation theory for the extended eigenvalue problem, where we keep all terms up to fourth order in $t_0$ and second order in $\varepsilon$. For the Fourier transform of the hopping term we find

$$H'_n = -t_0 J_n(\varepsilon)\sum_\sigma \left(c_{2\sigma}^+ c_{1\sigma} + c_{1\sigma}^+ c_{0\sigma}\right) + (-1)^n \times h.c., \qquad (23)$$

where $J_n(x) = \frac{1}{2\pi}\int_{-\pi}^{\pi} ds\, e^{i[x\sin(s)-ns]}$ is the $n$-th order Bessel function. The term $n = 0$ thus describes a reduction of the hopping with the factor $J_0(\varepsilon) = 1 - \varepsilon^2/4 + O(\varepsilon^2)$. At the extremely large amplitude $\varepsilon \approx 2.405$, the effective hopping vanishes which is known as coherent destruction of tunneling [20-21]. The terms $H'_n$ with $n \neq 0$ couple to higher Floquet sectors, which corresponds to a dressing of the levels with virtual absorption/emission of $n$ photons. Because $J_n(\varepsilon) \sim \varepsilon^n$ for $\varepsilon \to 0$, we can restrict ourselves to $n = \pm 1$ ($H'_{n\neq 0}$ does not contribute in first order perturbation theory). Summing up all hopping processes in $H'_{\pm 1}$ and $H'_0$ of the perturbation theory yields the result of the main text,

$$\Delta J = \frac{\varepsilon^2 t_0^4}{2} \left( \sum_{\pm} \left[ \frac{1}{U_1 \pm \hbar\omega} + \frac{1}{U_1} \right]^2 \frac{1}{U \pm \hbar\omega} - \frac{4}{U_1^2 U} - \frac{4}{U_1^3} \right). \qquad (24)$$

Exchange can be strengthened for low frequencies due to the effective lowering of the charge transfer energy by a virtual photon, and weakened for very high frequencies, where the coupling to higher Floquet bands is irrelevant and the coherent destruction of tunneling dominates. In experiments we typically have $\hbar\omega \sim U_1/2$, hence we anticipate a strengthening of the exchange interaction by the photon-assisted charge-transfer excitations under experimental conditions. We also note that in the strict limit $\omega \to 0$ (which is not relevant for experiments since it requires very long laser pulses) a small negative change $\Delta J$ appears when $\Delta > 0$.

Let us remark that Floquet theory makes predictions that will not be further discussed in the current manuscript. For example, the expression Eq. (24) shows that the effect of the electric field on the exchange interaction is strongly enhanced close to the resonance (where in a solid, however, one has strong absorption). In addition, different effects may occur in the non-perturbative regime, where the coupling to Floquet higher sectors is not negligible. Finally, the current model is clearly just a minimal model for laser-controlled super-exchange. A much richer behavior can be expected if more orbitals are included in the description.

*Results*

Before analyzing the effect of an electric field out of equilibrium, we study in equilibrium the quality of the general formulas Eq.(8) and Eq.(16), as well as the perturbative analytical expression Eq. (18) against the exact singlet-triplet splitting of the 3-site cluster. The result is shown in Fig. S10, where the different calculations of the exchange interaction are plotted as functions of $U/t_0$ for $\Delta/t_0 = 0.5$ and low temperature $\beta t_0 = 8000$, $\beta = 1/k_B T$.

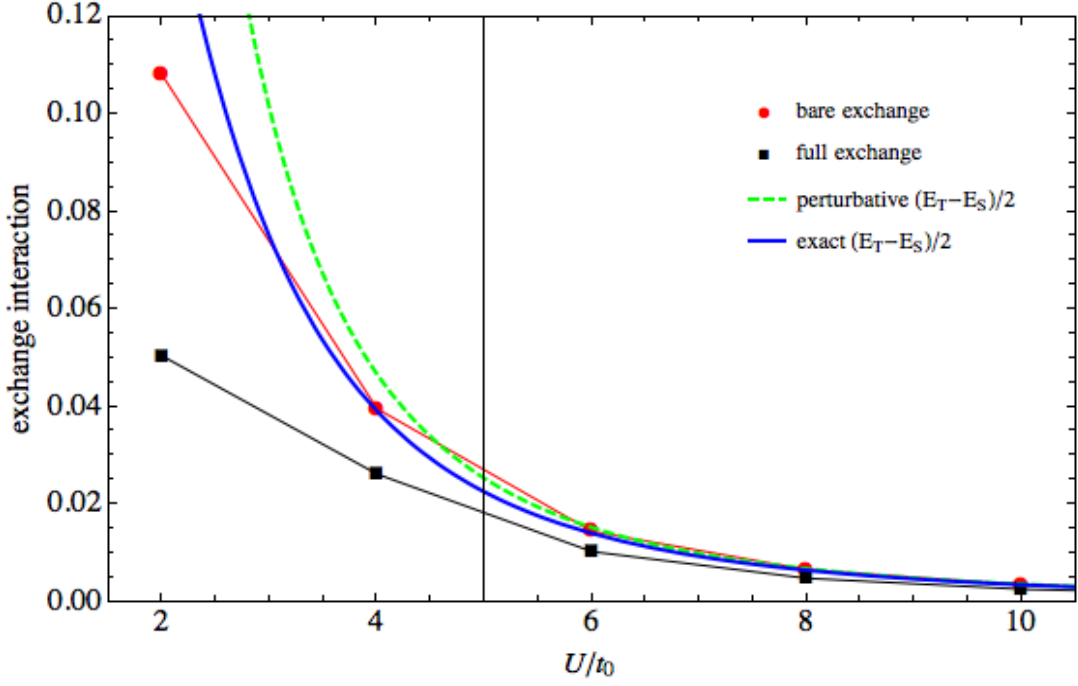

Fig. S10. Comparison of different expressions for the equilibrium exchange interaction as function of $U/t_0$.

It is observed that the full exchange interaction ($j_{02}^0$, black squares) introduces a systematic underestimation compared to the exact singlet-triplet splitting (blue solid lines) for all values of $U$. Instead, the bare exchange interaction ($J_{02}^0$, red discs) shows excellent numerical agreement already for relatively small $U/t_0 = 4$. We attribute the deviations at smaller $U$ to the neglect of vertex corrections in the general formulas. The perturbative analytical expression Eq. (24) (green dashed line) is quite accurate already at $U/t_0 = 6$.

Fig. S11 shows the relative change of the exchange interaction $\Delta J/J$ in the periodically driven cluster model as function of $\varepsilon^2$. For the numerical evaluation of the general formulas we slowly switched on the electric field using a rise time of 10 oscillation periods of the electric field pulse, using the model parameters $U/t_0 = 6$, $\Delta/t_0 = 0.5$, $\hbar\omega/t_0 = 3$, $\beta t_0 = 2000$. The results show an excellent quantitative agreement between the numerical results of the general formula, $J_{ab}^\varepsilon$ from Eq. (17), and the analytical Floquet theory Eq.(24), demonstrating an enhancement of the exchange interaction that scales linear with the intensity of the electric field. The order of magnitude of the effect in absolute numbers is discussed in the main text.

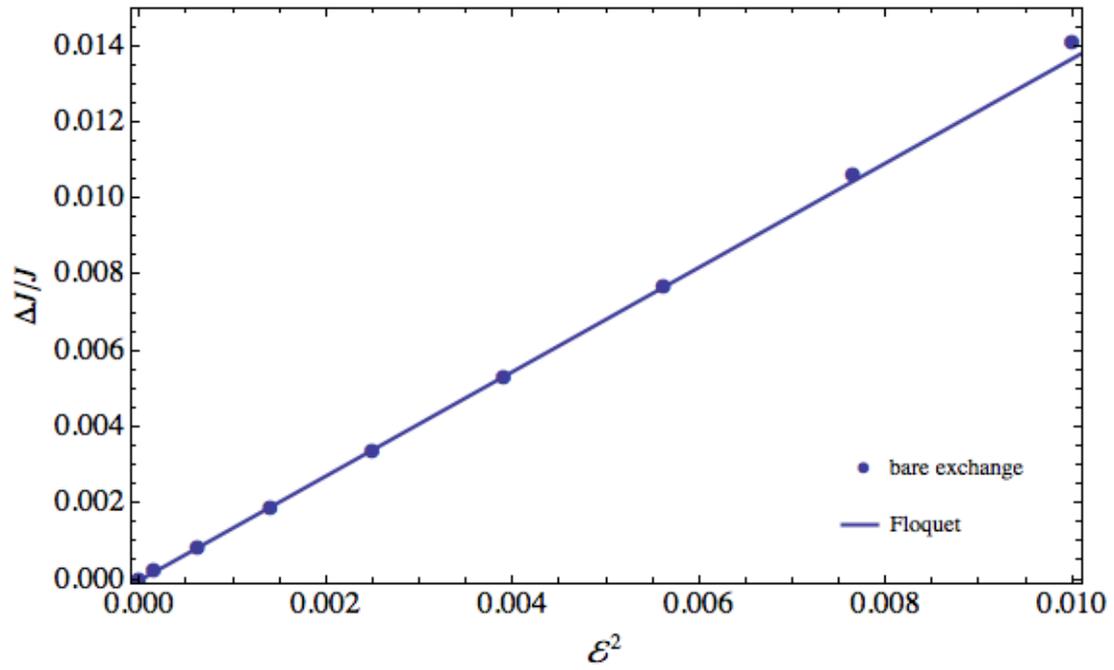

Fig. S11. Dependence of the exchange interaction on the applied electric field from the general formulas (blue dots) and the analytical Floquet theory (solid line).

**Supplementary note 7 – Macroscopic theory of the quasi-antiferromagnetic mode excitation via optical perturbation of the exchange interaction**

The equilibrium orientation of the iron spins in canted antiferromagnets is given by the minimum of the thermodynamic potential[2]:

$$\Phi = J\mathbf{S}_1\mathbf{S}_2 + 2\mathbf{D}\cdot[\mathbf{S}_1 \times \mathbf{S}_2] + K_x(S_{1x}^2 + S_{2x}^2) + K_y(S_{1y}^2 + S_{2y}^2) + K_z(S_{1z}^2 + S_{2z}^2) +$$
$$+ K_4(S_{1x}^4 + S_{1y}^4 + S_{1z}^4 + S_{2x}^4 + S_{2y}^4 + S_{2z}^4),$$

(25)

where $\mathbf{S}_1$ and $\mathbf{S}_2$ are the vectors that characterize the spins of the iron ions in the two magnetic sublattices, $J$ is the nearest neighbor isotropic exchange interaction constant; $\mathbf{D}$ is a constant vector pointing along the $y$-axis and describing the Dzyaloshinskii-Moriya antisymmetric exchange interaction; $K_x$, $K_y$, $K_z$, $K_4$ are the constants of the effective anisotropy. The $K_x$, $K_y$, $K_z$, $K_4$ are purely phenomenological and "effective" parameters in a sense, that they do not necessarily have a clear physical meaning, being a combinations of the single ion anisotropy terms and "hidden" exchange coupling parameters (see Ref. 22 for a detailed discussion).

It is instructive to rewrite the potential (25) in terms of the ferromagnetic vector (magnetization) $\mathbf{M} = -\gamma(\mathbf{S}_1 + \mathbf{S}_2)$ and the antiferromagnetic vector $\mathbf{L} = -\gamma(\mathbf{S}_1 - \mathbf{S}_2)$ as follows [1]

$$\Phi = -JM_{Fe^{3+}}^2 + \frac{1}{2}JM^2 + D[M_xL_z - M_zL_x] + \frac{1}{2}(K_y - K_x)L_y^2 + \frac{1}{2}(K_z - K_x)L_z^2 + \frac{1}{4}K_4L^4. \quad (26)$$

Here $\gamma$ is the absolute value of the gyromagnetic ratio of an electron and $M_{Fe^{3+}}$ is the magnetic moment of the iron ion. Note that $M^2 + L^2 = 4M_{Fe^{3+}}^2$. Let us consider ErFeO$_3$ or YFeO$_3$ as an example (the case of other materials is identical after rotation of the coordinate system by 90°). Taking into account that in canted antiferromagnets $J \gg D \gg K_x, K_y, K_z \gg K_4$ [1], the equilibrium magnetic configuration reads

$$M_x = M_y = L_y = L_z = 0, \ L_x = L_0 \approx 2M_{Fe^{3+}}, \ M_z = M_0 \approx \frac{D}{J}L_0. \quad (27)$$

Here $D/J$ defines the canting angle which is a small parameter and in the subsequent derivation all terms smaller than the canting angle are neglected.

The spin dynamics is described by the Landau-Lifshitz equations for ferromagnetic and antiferromagnetic vectors

---

[2]The formulae in this file are written in the Gaussian system of units, but the final answers are converted to SI units.

$$\frac{d\mathbf{M}}{dt} = \gamma\left(\left[\mathbf{M} \times \frac{\partial \Phi}{\partial \mathbf{M}}\right] + \left[\mathbf{L} \times \frac{\partial \Phi}{\partial \mathbf{L}}\right]\right), \quad \frac{d\mathbf{L}}{dt} = \gamma\left(\left[\mathbf{M} \times \frac{\partial \Phi}{\partial \mathbf{L}}\right] + \left[\mathbf{L} \times \frac{\partial \Phi}{\partial \mathbf{M}}\right]\right). \tag{28}$$

The vectors $\mathbf{M}$ and $\mathbf{L}$ can be represented as a sum of the equilibrium $\mathbf{M}_0$ and $\mathbf{L}_0$ and the time-varying $\mathbf{m}(t)$ and $\mathbf{l}(t)$ components. In the case of small deviations from equilibrium ($m << M_0$ and $l << L_0$) one can obtain a linearized system of equations describing the quasi-antiferromagnetic resonance:

$$\begin{cases} \dfrac{dm_z}{dt} = \gamma\left(K_y - K_x\right)L_0 l_y, \\[2mm] \dfrac{dl_x}{dt} = -\gamma\left(K_y - K_x\right)M_0 l_y, \\[2mm] \dfrac{dl_y}{dt} = -\gamma\left(JL_0 + DM_0\right)m_z + \gamma D L_0 l_x. \end{cases} \tag{29}$$

From the system (29) it follows that $l_y$ obeys the harmonic oscillator equation

$$\frac{d^2 l_y}{dt^2} + \omega_{\mathrm{qAF}}^2 l_y = 0, \tag{30}$$

where $\omega_{\mathrm{qAF}} \approx \gamma\sqrt{2 H_{\mathrm{E}} H_{\mathrm{A}}}$ is the frequency of the quasi-antiferromagnetic mode, $H_{\mathrm{E}} = \frac{1}{2}JL_0$ is the exchange field and $H_{\mathrm{A}} = \left(K_y - K_x\right)L_0$ is the anisotropy field.

Eqs. (29) and (30) describe the free dynamics of the spins without damping. To include the light-induced stimulus due to the inverse isotropic magnetic refraction, one has to consider an additional contribution, proportional to the envelope of the optical intensity $I_{\mathrm{opt}}$, $\Phi_{\mathrm{IMR}} = -\sum_{i,j} a I_{\mathrm{opt}} M^2$, added to the thermodynamical potential (25). As discussed above this term can be considered as a light-induced perturbation of the symmetric exchange energy, i.e.

$$\Phi_{\mathrm{ex}} = \Phi_{\mathrm{ex}\,0} + \Phi_{\mathrm{IMR}} = \frac{1}{2}J_0 M^2 - a I_{\mathrm{opt}} M^2 = \frac{1}{2}\left(J_0 + \Delta J\right)M^2, \tag{31}$$

where $\Delta J = -2 a I_{\mathrm{opt}}$, and the coefficient $a$ can be calculated from the microscopic theory of non-equilibrium exchange presented in the previous section. According to Eq. (24) the perturbation of the exchange $\Delta J(t)$ follows the envelope of the optical intensity, i.e. $\Delta J(t) = \Delta J_0 \exp\left(-t^2 / \tau^2\right)$, where $\Delta J_0$ is the peak change of the exchange parameter, $\tau = \tau_{\mathrm{FWHM}} / 2\sqrt{\ln 2}$, $\tau_{\mathrm{FWHM}}$ being the full width at half maximum of the laser pulse intensity envelope. However, it is also possible that the antisymmetric Dzyaloshinskii-Moriya energy $\Phi_D$ is also modified such as

$$\Phi_D = (D_0 + \Delta D)[M_x L_z - M_z L_x], \tag{32}$$

where $\Delta D$ is proportional to the envelope of the optical intensity, i.e. $\Delta D(t) = \Delta D_0 \exp(-t^2 / \tau^2)$. The light-induced modification of the exchange energy creates the torque $\sim \left[ \mathbf{L} \times \dfrac{\partial \Phi}{\partial \mathbf{M}} \right] + \left[ \mathbf{M} \times \dfrac{\partial \Phi}{\partial \mathbf{L}} \right]$ acting on the antiferromagnetic vector $\mathbf{L}$. It appears on the right-hand side of the equation of motion of $l_y$ as a driving force:

$$\frac{d^2 l_y}{dt^2} + \omega_{qAF}^2 l_y = -\gamma L_0 M_0 \frac{d(\Delta J)}{dt} + \gamma (L_0^2 - M_0^2) \frac{d(\Delta D)}{dt}. \tag{33}$$

Note, that here we restrict ourselves to the terms of first order in smallness with respect to $\Delta J$, $\Delta D$, $l_y$, $l_x$ and $m_z$. Moreover, since $L_0^2 >> M_0^2$ we can neglect the last term on the right side of Eq. (33). Let us also introduce some phenomenological damping into the equation of motion (33) as

$$\frac{d^2 l_y}{dt^2} + 2\nu \frac{dl_y}{dt} + \omega_{qAF}^2 l_y = -\gamma L_0 M_0 \frac{d(\Delta J)}{dt} + \gamma L_0^2 \frac{d(\Delta D)}{dt}, \tag{34}$$

where $\nu$ is the damping parameter.

Applying the Fourier transformation to Eq. (34) with respect to time $t$ we get

$$\omega^2 \tilde{l}_y + 2i\nu\omega\tilde{l}_y + \omega_{qAF}^2 \tilde{l}_y = \gamma L_0 (M_0 \Delta J_0 - L_0 \Delta D_0) \sqrt{\pi} \, i\,\omega\tau \exp\left(-\frac{\omega^2 \tau^2}{4}\right), \tag{35}$$

where $\tilde{l}_y(\omega)$ is a Fourier transform of $l_y(t)$

$$l_y(t) = \frac{1}{2\pi} \int_{-\infty}^{\infty} l_y(\omega) \mathrm{e}^{i\omega t} d\omega. \tag{36}$$

By rearranging Eq. (35) we obtain

$$\tilde{l}_y = -\gamma L_0 (M_0 \Delta J_0 - L_0 \Delta D_0) \frac{i\omega}{\omega_{qAF}^2 - \omega^2 + 2i\nu\omega} \sqrt{\pi}\tau \exp\left(-\frac{\omega^2 \tau^2}{4}\right). \tag{37}$$

At the same time, it follows from Eqs. (29) that $\mathrm{i}\,\omega\tilde{m}_z = \gamma L_0\left(K_y - K_x\right)\tilde{l}_y$, where $\tilde{m}_z$ is a Fourier transform of $m_z$. Thus, using Eq. (37) and the relations $\gamma^2 L_0^2\left(K_y - K_x\right) = \dfrac{\omega_{\mathrm{qAF}}^2}{J_0}$ and $\dfrac{M_0}{L_0} = \dfrac{D_0}{J_0}$ we can write

$$\tilde{m}_z = -\left(\frac{\Delta J_0}{J_0} - \frac{\Delta D_0}{D_0}\right)\omega_{\mathrm{qAF}}^2 M_0 \frac{1}{\omega_{\mathrm{qAF}}^2 - \omega^2 + 2\mathrm{i}\nu\omega}\sqrt{\pi}\,\tau \exp\left(-\frac{\omega^2\tau^2}{4}\right). \qquad (38)$$

It is instructive to consider the limit case of the instantaneous torque, i.e. $\tau\,\omega_{\mathrm{qAF}} \ll 2\pi$. In this case Eq. (38) can be approximated as

$$\tilde{m}_z \approx -\left(\frac{\Delta J_0}{J_0} - \frac{\Delta D_0}{D_0}\right)\omega_{\mathrm{qAF}}^2 M_0 \frac{1}{\omega_{\mathrm{qAF}}^2 - \omega^2 + 2\mathrm{i}\nu\omega}\sqrt{\pi}\,\tau \ . \qquad (39)$$

By calculating the inverse Fourier transform of Eq. (38) we can get an analytical solution for the magnetization response (in the case of small damping $\nu \ll \omega_{\mathrm{qAF}}$)

$$m_z(t) = \Theta(t)m_0 \sin\left(\omega_{\mathrm{qAF}}t\right)e^{-\nu t}, \qquad (40)$$

where $\Theta(t)$ is Heaviside function and

$$m_0 = -\left(\frac{\Delta J_0}{J_0} - \frac{\Delta D_0}{D_0}\right)\sqrt{\pi}\,\omega_{\mathrm{qAF}}\tau M_0 . \qquad (41)$$

Eq. (41) can be rewritten in the form

$$m_0 = -\left[\Delta\left(\frac{D}{J}\right)\Big/\left(\frac{D_0}{J_0}\right)\right]\sqrt{\pi}\,\omega_{\mathrm{qAF}}\tau M_0 , \qquad (42)$$

where.

$$\Delta\left(\frac{D}{J}\right) = \frac{D_0 - \Delta D_0}{J_0 - \Delta J_0} - \frac{D_0}{J_0} . \qquad (43)$$

Eq. (42) shows that the ratio between the magnetization deviation from equilibrium $m_0$ and the spontaneous magnetization $M_0$ is directly proportional to the relative change of the ratio $\dfrac{D}{J}$.

**Supplementary note 8 – Electrodynamics of THz generation in canted ferromagnets**

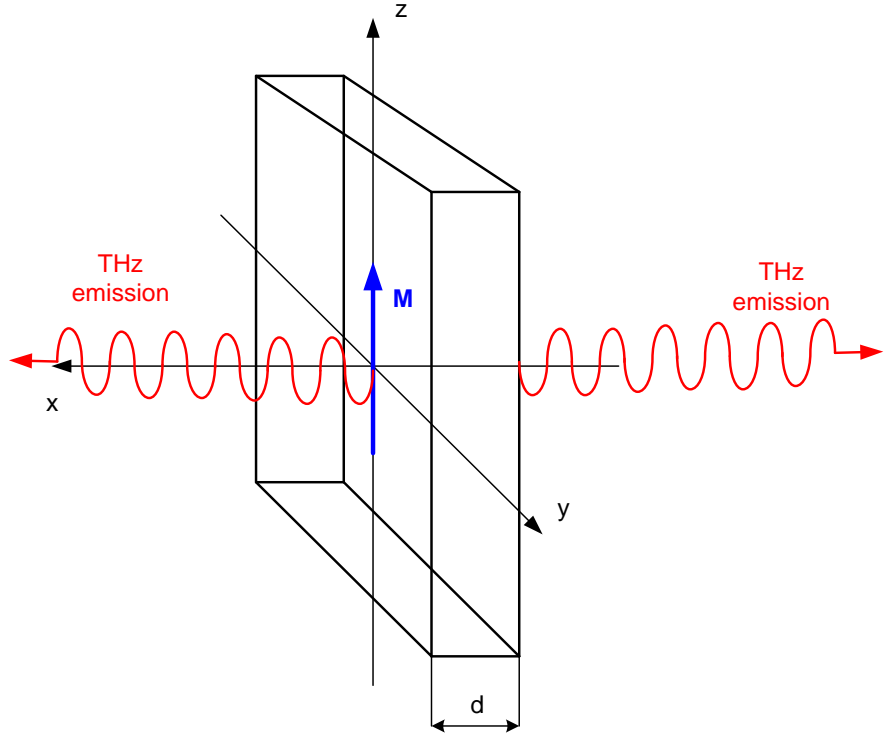

Fig. S12. Geometry of THz generation in a slab of a canted antiferromagnet.

The optical pulse excites the magnetization oscillations in a thin slab of a canted antiferromagnet. The lateral dimensions of the sample and the size of the excitation spot (several mm) are much larger than the THz wavelength (300 μm). Thus we can restrict ourselves to a one-dimensional problem as follows.

Let us consider an infinite slab of a material of thickness $d$ (see Fig. 13) with the permittivity $\varepsilon_s$, containing an oscillating magnetization $\mathbf{M}$ oriented along the $z$-axis in the form

$$\mathbf{M}(t,x) = (M_0 + m(t))[\Theta(x+d) - \Theta(x)]\mathbf{z}_0 \qquad (44)$$

where $M_0$ is the equilibrium magnetization, $m(\text{t})$ is the time-dependent deviation from equilibrium, and $\Theta(x)$ is a Heaviside function. The magnetic permeability in the canted antiferromagnets in the vicinity of the magnetic resonance frequency varies from 0.9 to 1.1 [1], and thus can be neglected. From the physical point of view, it means that the inverse action of the emitted magnetic field on the oscillating magnetization is negligible.

To find the emission of the magnetization (44) one has to solve the wave equation for the $y$-component of the electric field $E_y$

$$\frac{\partial^2 E_y}{\partial x^2} - \frac{\varepsilon}{c^2}\frac{\partial^2 E_y}{\partial t^2} = \frac{4\pi}{c^2}\frac{\partial j}{\partial t}, \tag{45}$$

where $j = c\dfrac{\partial M}{\partial x} = cm(t)[\delta(x+d) - \delta(x)]$, $\delta(x)$ is a Dirac function and the permittivity $\varepsilon(x)$ in the THz range is $\varepsilon_s$ in the slab ($-d < x < 0$) and unity outside, respectively. Eq. (45) is derived from the Maxwell equations.

After applying the Fourier transformation with respect to time, Eq. (45) transforms to

$$\frac{\partial^2 \tilde{E}}{\partial x^2} + \frac{\varepsilon(x)\omega^2}{c^2}\tilde{E} = \frac{4\pi}{c}\,\mathrm{i}\,\omega\tilde{m}(\omega)[\delta(x) - \delta(x+d)]. \tag{46}$$

We solve Eq. (46) in the homogeneous regions $x < -d$, $-d < x < 0$ and $x > 0$ and match the solutions by the boundary conditions that arise after integrating Eq. (46) across the boundary at $x = -d$ and $x = 0$. These boundary conditions imply the continuity of $\tilde{E}$, while $\dfrac{\partial \tilde{E}}{\partial x}$ exhibits a finite discontinuity of $\dfrac{4\pi}{c}\,i\omega\tilde{m}(\omega)$ and $-\dfrac{4\pi}{c}\,i\omega\tilde{m}(\omega)$ at the boundaries $x = -d$ and $x = 0$ respectively. The solution has the form

$$\tilde{E} = \begin{cases} R\,\mathrm{e}^{-\mathrm{i}k_0 x}, & x > 0 \\ F\,e^{-\mathrm{i}kx} + B e^{\mathrm{i}kx}, & d < x < 0 \\ C\,\mathrm{e}^{\mathrm{i}k_0(x+d)}, & x < d \end{cases}, \tag{47}$$

where $k_0 = \dfrac{\omega}{c}$ is the wavenumber in free space, $k = \dfrac{\omega}{c}\sqrt{\varepsilon_s} = \dfrac{\omega}{c}n$ is the wavenumber in the slab with refractive index $n$. In our experiment we measured the field $E(t)$ emitted into free space given by

$$E(t) = \frac{1}{2\pi}\int_{-\infty}^{\infty} R(\omega)\mathrm{e}^{\mathrm{i}\omega t}\,d\omega. \tag{48}$$

where

$$R(\omega) = \frac{4\pi\omega}{c}\,\tilde{m}(\omega)\frac{2k + (k_0 - k)e^{-\mathrm{i}kd} - (k_0 + k)e^{\mathrm{i}kd}}{(k-k_0)^2 e^{-\mathrm{i}kd} - (k-k_0)^2 e^{\mathrm{i}kd}}. \tag{49}$$

The function $\tilde{m}(\omega)$ has been already derived above and given by Eq. (38).

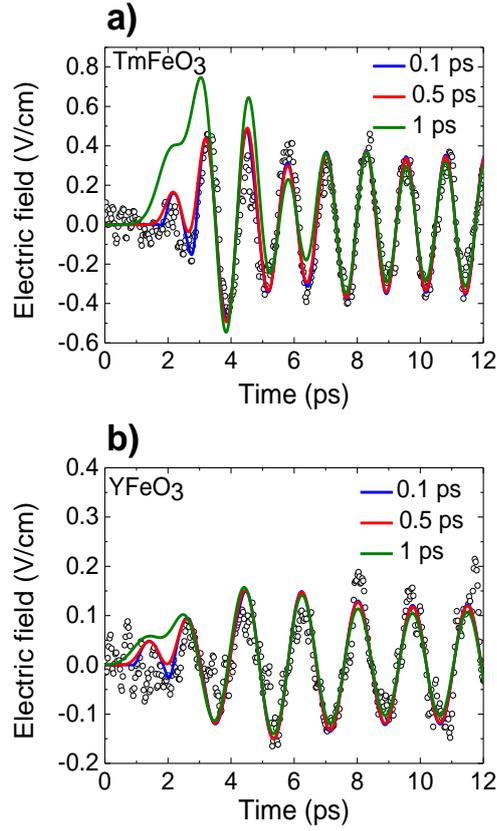

Fig. S13. Comparison of the experimental signals with the waveforms calculated theoretically. The signal generated in TmFeO$_3$ at 50 K (open circles) shown together with the theoretical waveforms (solid curves) calculated assuming different durations of the Gaussian torque (a). The signal generated in YFeO$_3$ at 15 K (open circles) shown together with the waveforms calculated theoretically assuming different durations of the torque (b).

The refractive index at the quasi-antiferromagnetic frequency equals $n = 4.71$ and 4.68 for TmFeO$_3$ and YFeO$_3$, respectively [1]. The thickness of our samples equals $d = 60$ μm and 70 μm correspondingly. By substituting these parameters into Eqs. (48) and (49) and fitting the amplitude, the time delay corresponding to the arrival of the pump pulse and the damping constant in Eq. (38) the waveforms shown in Fig. S13 were obtained for different excitation torque durations. It is seen in the figure that the waveforms calculated for the excitation torque with a duration of < 0.5 ps agree well with the experimental data. As the duration of the torque exceeds 0.5 ps the calculated waveforms become different from the experimental signals because the spins start to follow the torque adiabatically, leading to the generation of a broadband electric field pulse not observed in the measurements. When the rise-time becomes longer than 1 ps the excitation of the quasi-antiferromagnetic mode gradually vanishes since the spectrum of the torque does not overlap with the resonance frequency anymore. Importantly, to keep the amplitude of the excited quasi-antiferromagnetic oscillation of the same amplitude one has to increase the peak amplitude of the torque as its duration becomes longer.

However, even for the case of the shortest torque (100 fs) both calculated and measured signals demonstrate a finite rise-time($\sim$ 1 ps). This effect is due to the fact that the samples act as Fabry-Perot resonators for the electromagnetic radiation at the frequencies of interest. The Fabry-Perot behavior is described by the term $\dfrac{2k + (k_0 - k)e^{-ikd} - (k_0 + k)e^{ikd}}{(k - k_0)^2 e^{-ikd} - (k - k_0)^2 e^{ikd}}$ in Eq. (49). The electric field emitted by the instantaneously commencing magnetization oscillation does not follow the magnetization immediately but exhibits a finite rise time determined by the characteristic timescale of the resonator $\dfrac{nd}{c} \sim 1$ ps. The presence of Fabry-Perot resonances in the vicinity of quasi-antiferromagnetic frequency was also experimentally verified in our THz absorption measurements.

In FeBO$_3$ and $\alpha$-Fe$_2$O$_3$ the period of quasi-antiferromagnetioc oscillation is several times smaller than in the orthoferrites. Therefore, excitation is possible for longer torques. However, it is natural to assume that the mechanism of excitation and its timescale is the same in these materials as in the orthoferrites.

Using Eq. (49) and measured values of the electric field we estimated the amplitude of the oscillating magnetization as $\sim 10^{-3}$emu/cm$^3$in the orthoferrites and the iron borate ($\sim$1 A/m in SI units), which implies $\Delta\left(\dfrac{D}{A}\right) / \left(\dfrac{D_0}{A_0}\right) \approx 0.01$ %. The amplitude of the electric field generated in the hematite is smaller, but this material absorbs the light at 800 nm that leads to a different regime of the excitation. Thus, the radiated emission might have been generated in a thin surface layer in which the optical pulse penetrates, leading to a similar efficiency of the exchange energy perturbation.

**Supplementary note 9 – The strength of the optical control of the super-exchange**

Even a relatively small change of the exchange interaction corresponds to rather large effects expressed in absolute units of equivalent field and energy.

From Eq. (33) one can see that the changing of the ratio $D/J$ is equivalent to the application of the short pulsed magnetic field $\mathbf{b}_{\text{eff}}$ along the magnetization direction. The peak amplitude of this field is

$$b_0 = \left( \Delta J_0 M_0 - \Delta D_0 L_0 \right) = \frac{J_0 m_0}{\sqrt{\pi} \, \omega_{\text{qAF}} \tau} \,. \qquad (50)$$

For the parameters of our experiment and the known strength of the exchange field in the materials under study ($J_0 L_0 \sim 1000$ Tesla) we estimate the amplitude of the equivalent field to be of the order of $\sim 0.01$ Tesla per 1 mJ/cm$^2$ pump fluence. This value is of the same order of magnitude as the maximal strength of the light-induced magnetic field (per the same pump fluence) achieved with help of opto-magnetic phenomena such as the inverse Faraday effect [23]. However, the inverse Faraday effect and similar phenomena owe their strength to the spin-orbit coupling and therefore are not strong in all magnetic materials. Therefore the exchange driven optical control of spins should be a more versatile tool for the manipulation of magnetic states. The short single-cycle THz pulses of magnetic field which are shown to excite spin dynamics can achieve a peak amplitude of 0.1 Tesla [24]. Unfortunately, to generate such a pulse one needs extremely high pump fluences and dedicated laser systems not widely available. Indeed, the pulses of magnetic field used in Ref. 24 were generated with a help of optical pulses with energy of 5 mJ, that is larger than the total energy of pulses generated in a normal amplified Ti:sapphire laser used in our measurements.

The energy of the interaction between light and magnetic system can be estimated as

$$\Delta W = V b_0 M_0 \,. \qquad (51)$$

where $V$ is the optically excited volume of the material ($\sim 100\ \mu\text{m} \times 1$ mm $\times 1$ mm in our measurements). Using Eq. (50) and taking $M_0 \approx 10$ emu/cm$^3$ we estimate the energy $\Delta W$ is of the order of 1 μJ per excited area of 1 cm$^2$.

Finally we note that the 0.01% change of the ratio of the two exchange constants represents a difference of the relative changes of each of them. This means that if the changes are of the same sign (which is quite likely) then each of them could in fact be much greater than 0.01% in agreement with the prediction of the microscopic theory.